\documentclass[fleqn,10pt]{wlscirep}
\usepackage[utf8]{inputenc}
\usepackage[T1]{fontenc}
\usepackage{soul,ulem}
\usepackage{lscape,longtable}
\usepackage{threeparttable}
\usepackage{endnotes}

\title{Understanding hesitancy with revealed preferences across COVID-19 vaccine types}

\author[1]{Kristóf Kutasi}
\author[2,3,4]{Júlia Koltai}
\author[5,6]{Ágnes Szabó-Morvai}
\author[7]{Gergely Röst}
\author[4,8]{M\'arton Karsai}
\author[9,10,+]{Péter Biró}
\author[11,12,+,*]{Balázs Lengyel}
\affil[1]{Rice University, Department of Economics, Houston TX, 77005-1827, USA}
\affil[2]{Centre for Social Sciences, Computational Social Science - Research Center for Educational and Network Studies, Budapest, 1097, Hungary}
\affil[3]{Eötvös Loránd University, Faculty of Social Sciences, Budapest, 1117, Hungary}
\affil[4]{Central European University, Department of Network and Data Science, Vienna, 1100, Austria}
\affil[5]{Eötvös Loránd Research Network, Centre for Economic and Regional Studies, Health and Population Lendület Research Group, Budapest, 1097, Hungary}
\affil[6]{Debrecen University, Department of Economics, Debrecen, 4032, Hungary}
\affil[7]{University of Szeged, Bolyai Institute, Szeged, 6722, Hungary}
\affil[8]{Alfréd Rényi Institute of Mathematics, Budapest, 1053, Hungary}
\affil[9]{Eötvös Loránd Research Network, Centre for Economic and Regional Studies, Mechanism Design Lendület Research Group, Budapest, 1097, Hungary}
\affil[10]{Corvinus University of Budapest, Department of Operations Research and Actuarial Sciences, Budapest, 1093, Hungary}
\affil[11]{Eötvös Loránd Research Network, Centre for Economic and Regional Studies, Agglomeration and Social Networks Lendület Research Group, Budapest, 1097, Hungary}
\affil[12]{Corvinus University of Budapest, Corvinus Institute for Advanced Studies, Budapest, 1093, Hungary}

\affil[*]{Corresponding author: lengyel.balazs@krtk.hu}
\affil[+]{these authors jointly supervised this work}

\begin{abstract}
Many countries have secured larger quantities of COVID-19 vaccines than their populace is willing to take. This abundance and variety of vaccines created a historical moment to understand vaccine hesitancy better. 
Never before were more types of vaccines available for an illness and the intensity of vaccine-related public discourse is unprecedented. Yet, the heterogeneity of hesitancy by vaccine types in certain segments of society has been neglected so far, even though factual or believed vaccine characteristics and patient attributes are known to influence acceptance.
In this paper, we address this problem by analysing acceptance and assessment of five vaccine types using information collected with a nationally representative survey (N=1000) at the end of the third wave of the COVID-19 pandemic in Hungary, where a unique portfolio of vaccines were available to the public in large quantities. Our special case enables us to quantify revealed  preferences across vaccine types since one could evaluate a vaccine unacceptable and even could reject an assigned vaccine to wait for another type. We find that the source of information that respondents trust characterizes their attitudes towards vaccine types differently and leads to divergent vaccine hesitancy. Believers of conspiracy theories were significantly more likely to evaluate the mRNA vaccines (Pfizer and Moderna) unacceptable while those who follow the advice of politicians evaluate vector-based (AstraZeneca and Sputnik) or whole-virus vaccines (Sinopharm) acceptable with higher likelihood. We illustrate that the rejection of non-desired and re-selection of preferred vaccines fragments the population by the mRNA versus other type of vaccines while it generally improves the assessment of the received vaccine. These results highlight that greater variance of available vaccine types and individual free choice are desirable conditions that can widen the acceptance of vaccines in societies.

\end{abstract}
\begin{document}

\flushbottom
\maketitle

\thispagestyle{empty}

\section{Introduction}

The fast development of vaccines against the SARS-COV-2 virus will go to history books as an outstanding scientific achievement \cite{cohen2020vaccine, graham2020rapid}. Unprecedented efforts and investments resulted vaccines of various types \cite{corey2020strategic}: established techniques yielded vector-based and whole-virus vaccines while cutting-edge technologies were pioneered for mRNA vaccines. However, mostly due to misinformation on vaccines, and uncertainties about their effects on health, 
large populations remained hesitant against vaccination \cite{lazarus2021global,machingaidze2021understanding, sallam2021covid, solis2021covid}, threatening the efficiency of vaccination plans \cite{wouters2021challenges, giordano2021modeling, bubar2021model, matrajt2021vaccine, priesemann2021towards}. 
Previous literature on vaccine hesitancy has considered the individual act of taking a vaccine an uncertain decision because risks and benefits are difficult to compare \cite{rosselli2016old}. Hence, besides demographic and socio-economic characteristics that correlate with hesitancy \cite{loomba2021measuring, paul2021attitudes, sonawane2021covid}, personal beliefs are thought to be critical as well\cite{troiano2021vaccine, freeman2021effects}. Consequently, the trusted source of information is an important factor: those who trust public authorities and scientists are generally less hesitant \cite{lazarus2021global, lindholt2021public, schernhammer2021correlates}, while believers of conspiracy theories are more hesitant about the vaccines\cite{salali2020covid, murphy2021psychological, soveri2021unwillingness}. 


Despite the intensive public discourse about the efficacy \cite{lipsitch2020understanding} and health risks of specific vaccines \cite{ledford2021could}, how the level of hesitancy differs across vaccine types is unclear. It makes vaccine comparison difficult that most countries apply only a few selected vaccines. Yet, individuals might not be hesitant against all types of vaccines; instead, might be open for vaccines that use certain technologies or come from certain countries \cite{schwarzinger2021covid}. However, such insights were gained so far only by investigating hypothetical vaccines \cite{dror2021vaccine}. Therefore, to increase vaccine acceptance, it is important to better understand how individual characteristics, and especially how the trusted source of information, can explain the hesitancy about particular vaccine types.


In this paper, we compare the acceptance and assessment of five vaccine types in Hungary. This country applied the greatest diversity of vaccines in the third wave of the COVID-19 pandemic\endnote{Vaccine diversity is simply calculated by the number of available vaccine types from data at https://ourworldindata.org/covid-vaccinations.}. Besides the centralized European Union (EU) purchase, which included the Pfizer-Biontech, Moderna, and AstraZeneca vaccines, the Hungarian government imported Sputnik from Russia and Sinopharm from China. (Although Janssen was also available in Hungary from May 2021, only a small share of population received it. We excluded this vaccine from the analysis.)\endnote{Official names of the vaccines are listed here. \textbf{Pfizer-Biontech}: Comirnaty/BNT162b2, COVID-19 mRNA vaccine. \textbf{Moderna}: Spikevax/mRNA-1273. \textbf{AstraZeneca}: Vaxzevria, Covishield / AZD1222, ChAdOx1 nCoV-19, ChAdOx1-S, AZD2816. \textbf{Sputnik}: Gam-COVID-Vac. Sinopharm: \textbf{Sinopharm}: BIBP COVID-19 vaccine / Zhong'aikewei, Hayat-Vax. \textbf{Janssen}: Ad26.COV2.S JNJ-78436735 Ad26COVS1 VAC31518} Although the application of Sputnik and Sinopharm put Hungary among the quickest countries globally in the early vaccination period and vaccination rate was still outstanding among Central and Eastern European countries at the beginning of the fourth wave\endnote{According to https://ourworldindata.org/covid-vaccinations, vaccination per capita was much higher in May 2021, when our data was collected, in Hungary than in other Central and Eastern European countries where trust in public authorities is at a similar level.}, the vaccination plan generated an intensive public debate. The government strongly communicated the safety and efficacy of Sputnik and Sinopharm \cite{goodwin2021psychological}; while mRNA vaccines were rated high in society \cite{koltai2021changing}. As hesitancy was high in the beginning of the vaccination campaign \cite{lindholt2021public} ($53\%$ in January 2021 and $20\%$ in May 2021 \cite{koltai2021changing}) a rigid vaccine assignment could have potentially decreased people's willingness to vaccinate. Thus, individuals were allowed to reject the assigned vaccine and wait for the preferred one. 

This historically unique situation enables us to contribute to the ongoing discussion in two ways. First, the globally outstanding diversity of available vaccines in a single country allows us to compare hesitancy across vaccines by technologies (mRNA, vector-based, and whole virus). We are able to identify the individual characteristics that correlate with hesitancy of specific vaccines, focusing on the role of trusted source of information. Second, the special allocation mechanism (i.e. the possibility to reject a vaccine and wait for another) allows us to analyze whether free choice across vaccine types can mitigate hesitancy.


Our empirical analysis is based on a nationally representative survey (N=1000) that we conducted at the end of the third wave of the pandemic when vaccines were already available for everyone\endnote{Number of cases can vary among analyses. For a precise consideration of sample sizes, please see Materials and Methods.} We collected detailed demographic, socio-economic, and health information of respondents. We asked respondents to identify sources they accept vaccine-related advice from (these include medical doctors, scientists, anti-vaccine propagators, politicians, family etc). They were also requested to evaluate each vaccine and clarify whether their assessment has changed over the third wave. Finally, a list of questions were used to map their COVID-19 infection and vaccination history including the date of potential positive tests, date and type of vaccines received and date and type of vaccines rejected (18\% of our vaccinated respondents rejected an assigned vaccine and later took another type). This latter information provides new insights into the nature of vaccine hesitancy through the analysis of revealed vaccine preferences 
and allows for the comparison of vaccine assessments between rejected and accepted vaccines. From now on, we use the word \(evaluate\) to refer on whether individuals considered the vaccine types acceptable; \(reject\) refers to the act when the individual was not willing to take the assigned vaccine; while \(accept\) is used to describe when a vaccine is taken.

We find heterogeneous vaccine preferences. Most individuals prefer mRNA vaccines but a non-negligible portion ($5.5\%$) of respondents who rejected a vaccine preferred vector-based and whole virus vaccines against mRNA types. The source of vaccine-related advice is an important factor of vaccine hesitancy. We find evidence for divergent attitudes towards vaccine types that can depend on the trusted source of advice. Believers of conspiracy theories are significantly more likely to evaluate the mRNA vaccines unacceptable. Those who follow the advice of national politicians are more likely to evaluate less popular vector-based and whole virus vaccines acceptable, even those that are not approved by the EU. 
We illustrate that the rejection of non-desired and re-selection of preferred vaccines generally improve the assessment of received vaccines. However, these revealed preferences fragments the population into groups of individuals that either accepts the newest but reject the more established vaccine technologies and vice versa. 

The remaining of the article has the following structure. We first introduce our data and illustrate patterns and dynamics of revealed preferences across vaccine types. Then, we analyse determinants of individual hesitancy by vaccine types using statistical techniques. Finally, we investigate vaccine assessment across those who have accepted or rejected the assigned vaccine. Our results call for diversified communication and vaccine allocation strategies to convince the hesitant population. Government communication can be effective in convincing the local population about less popular vaccines. However, misinformation must be fought to mitigate hesitancy towards the newly developed vaccines that might partly originate in uncertainties around new technologies. Although our analysis cannot prove cause and effect, the suggestive evidence can be used to develop a new hypothesis. In the Discussion, we argue that a greater variance of available vaccine types and individual free choice are desirable conditions that can widen the acceptance of vaccines in societies.

\section{Results}

\subsection{Vaccine assignment and revealed preferences}
Let us give a general overview about the logistics of the COVID-19 vaccination procedure in Hungary over the third wave of the COVID-19 pandemic, 2021 Spring. A vaccine seeking individual (patient) first had to register online at a government portal\endnote{https://vakcinainfo.gov.hu/regisztracio-oltasra}. The general physician (GP) received the list of patients in its region and contacted them in order of priority, based on chronic illness, age and occupation. Upon contacting a patient, the GP offered a type of vaccine to the patient. The offer followed the guidence of central vaccination plan and depended on patient characteristics (as explained above) and the availability of vaccines. However, the patient could either accept or reject the offered vaccine. In case of a rejection, the GP had to reach out to the patient later, when a different type of vaccine became available. This procedure characterized early vaccination until the vast majority of priority groups were vaccinated. Starting from April, when vaccines became increasingly available, patients could register and select available vaccines through a web-application (first AstraZeneca, then Sinopharm and Sputnik, and later other vaccines as well).

The vaccine allocation procedure in Hungary is especially interesting as patients could choose over the type of vaccination they would like to get. The individual's vaccine selection was in sharp contrast with the allocation mechanism used in other EU countries or in the United States in 2021 Spring where patients found out the assigned vaccine type only when they arrived to the vaccination center. Besides giving floor for various individual preferences, the Hungarian mechanism might have provided acceptable alternatives for those who got scared from the assigned vaccine. For example, the Sinopharm vaccine was assigned initially in Hungary to the older generation, despite it was recommended to a younger population. Similarly, AstraZeneca was assigned for the middle-aged generation, but many became afraid when the connection between the vaccine and thrombosis cases were discussed. 

\begin{table}[!b]
\centering
\caption{\label{tab:example}Percentage of individuals in the sample by major socio-demographic characteristics and COVID vaccination history. Percentages in the last three columns add up to 100$\%$ by character groupings (except the rows on specific vaccine types where the last column is not relevant). Individuals are classified by type of vaccine taken in the last rows.}
\begin{tabular}{lccccc}
  \hline
 & Evaluated all  & Evaluated one vaccine & Vaccinated & Vaccinated after & Non-vaccinated \\ 
 & vaccines  & acceptable but & without rejection & rejection &  \\ 
 &  unacceptable &  another unacceptable &  &  & \\
  & (\(\%\)) &  (\(\%\)) & (\(\%\)) &  (\(\%\)) & (\(\%\)) \\
  \hline
Men & 1.1 & 7.9 & 30.7 & 4.6 & 11.7 \\ 
  Women & 1.2 & 10.3 & 32.2 & 6.2 & 14.5 \\ 
    \hline
  University & 0.2 & 3.4 & 14.4 & 2.9 & 4.3 \\ 
  High-school & 1.7 & 11.7 & 32.7 & 6.2 & 14.7 \\ 
  Elementary & 0.4 & 3.1 & 15.9 & 1.7 & 7.2 \\ 
    \hline
  Not Chronic Illness & 1.6 & 11.8 & 33.0 & 6.4 & 19.9 \\ 
  Chronic Illness & 0.7 & 6.4 & 29.9 & 4.4 & 6.3 \\ 
  \hline
  Age 20-39 & 0.9 & 7.8 & 16.2 & 4.3 & 12.6 \\ 
  Age 40-59 & 0.8 & 5.9 & 20.6 & 3.4 & 9.2 \\ 
  Age 60-79 & 0.5 & 4.0 & 23.0 & 2.7 & 3.3 \\ 
  Age 80+ & 0.0 & 0.2 & 2.3 & 0.0 & 0.3 \\ 
    \hline
  Pfizer & - & 5.0 & 20.3 & 5.8 & - \\ 
  Moderna & - & 0.8 & 3.5 & 1.0 & - \\ 
  AstraZeneca & - & 1.1 & 12.1 & 1.1 & - \\ 
  Sputnik & - & 1.4 & 12.3 & 1.3 & - \\ 
  Sinopharm & - & 1.5 & 14.7 & 1.6 & - \\ 
   \hline
\end{tabular}
\label{survey_description}
\end{table}

To gain insights into these revealed individual preferences on vaccine types, we have conducted a nationally representative survey data collection on a sample of $1,000$ people with CATI (computer assisted telephone interviewing) \cite{karsai2020hungary,koltai2021monitoring} between May 25 and 31, 2021. Until this date the majority of non-hesitant population has already received at least one dose of vaccine \cite{koltai2021changing}. The data is representative for the adult Hungarian population by gender, age, education and domicile. For more detailed description of the data collection and survey methods see the Methods Section.

Table \ref{survey_description} describes the vaccination related variables by socio-demographic characteristics of the respondents. Evaluation and rejection of vaccines - which information makes our dataset unique - are presented by sex, education level, chronic illness, and age categories.
Here we can distinguish those who rate all vaccine types unacceptable ($2.3\%$) from those who evaluate at least one vaccine unacceptable but evaluate another acceptable ($18.2\%$). Further, we can compare those who accepted the first vaccine offered ($62.9\%$) with those who rejected the first offer ($10.8\%$). $26.3\%$ of the sample has not received any vaccine until the data collection. The ratio of non-vaccinated individuals in the total population of the country was around $32\%$ in June 2021. This moderate bias might be due to the tendency that those who are more concerned about the pandemic situation are more willing to take vaccines and participate in related surveys, or, to conform to the public norm, some unvaccinated might not have been admitted it. Apparently, women, the younger cohorts, the lower educated and those with no chronic disease are more likely to reject an offer or a vaccine in general.

\begin{figure}[hb]
\centering
\includegraphics[width=0.7\linewidth]{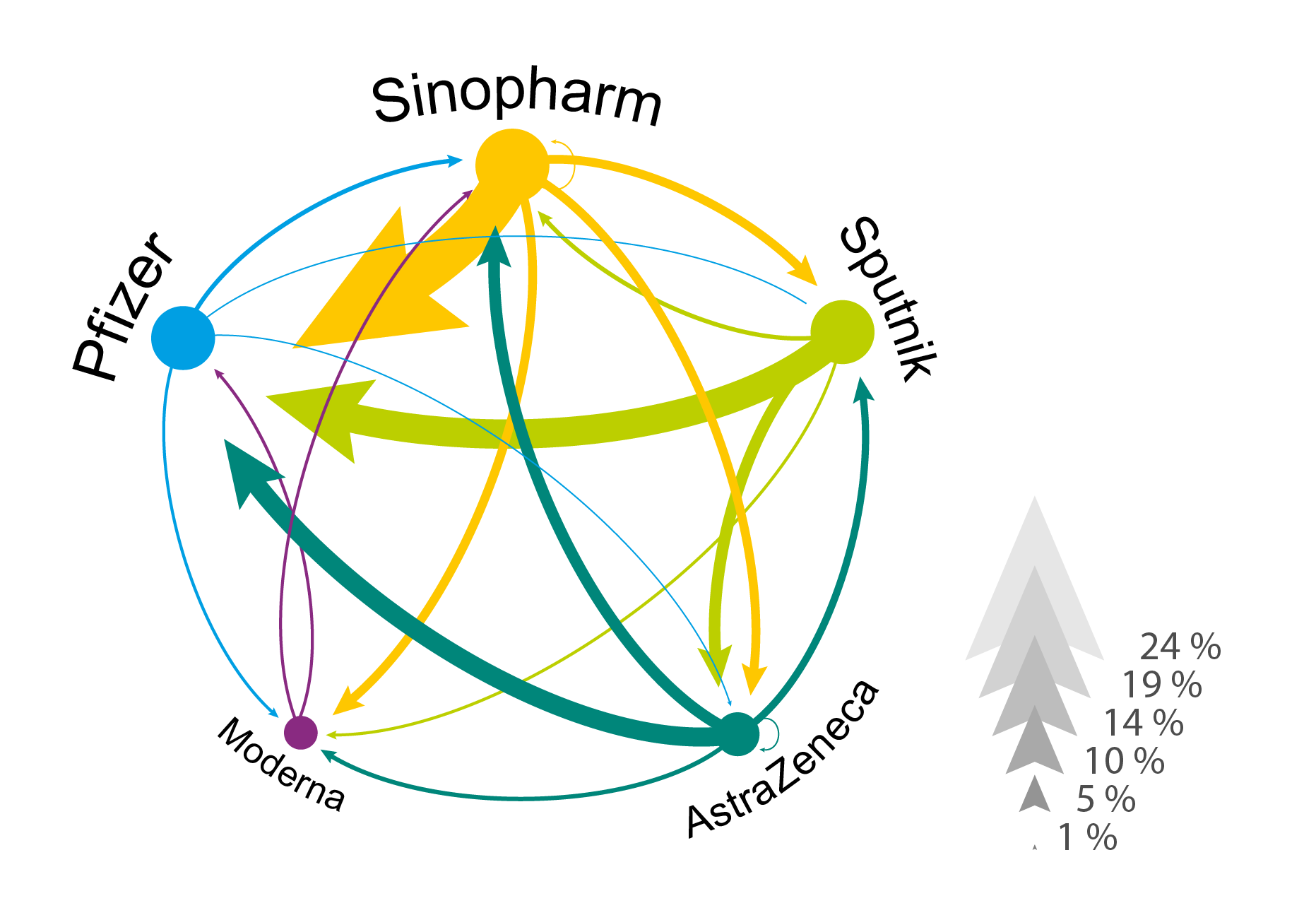}
\caption{\textbf{Revealed vaccine preferences by rejection and re-selection.} The network of revealed vaccine preferences in Hungary. Vaccines are linked if someone in our data has rejected one vaccine and later accepted another. The direction of the link goes from rejected to accepted and the width of the arrow corresponds to the percentage of individuals who acted accordingly among those who rejected a vaccine.}
\label{fig:Fig1}
\end{figure}

The data allow us to investigate acceptance and revealed preferences by vaccine types. Table \ref{survey_description} demonstrates that $5\%$ of those respondents who have received Pfizer rejected at least one other vaccine; while these shares are lower in the case of those who received other vaccine types. A similar pattern characterizes our sample according to the rejection of vaccines: $20.3\%$ of all respondents have accepted Pfizer without rejecting it while $5.8\%$ waited for Pfizer after rejecting another. There are fewer individuals in the data who accepted other assigned vaccines without rejection but except Moderna ($3.5$), which had the shortest supply in the country, AstraZeneca ($12.1\%$), Sputnik ($12.3\%$), and Sinopharm ($14.7\%$) were accepted by large populations without hesitancy. Although a smaller fraction of society than in the case of Pfizer, but few individuals waited for these latter vaccines specifically.

Figure \ref{fig:Fig1} illustrates the network of revealed preferences across vaccine types from our data. We observe a high ratio of individuals, among those who rejected a vaccine, who received Pfizer but rejected Sinopharm ($23\%$), Sputnik ($15\%$) or AstraZeneca ($9.7\%$). However, individuals also accepted these latter vaccines after rejecting other types. We see a few individuals who rejected the most accepted and then accepted a less popular vaccine. For example, $2.2\%$ of the hesitant population have rejected Pfizer to receive Sinopharm.  The rules of network generation is described in the Methods section.

\begin{figure}[!t]
\centering
\includegraphics[width=\linewidth]{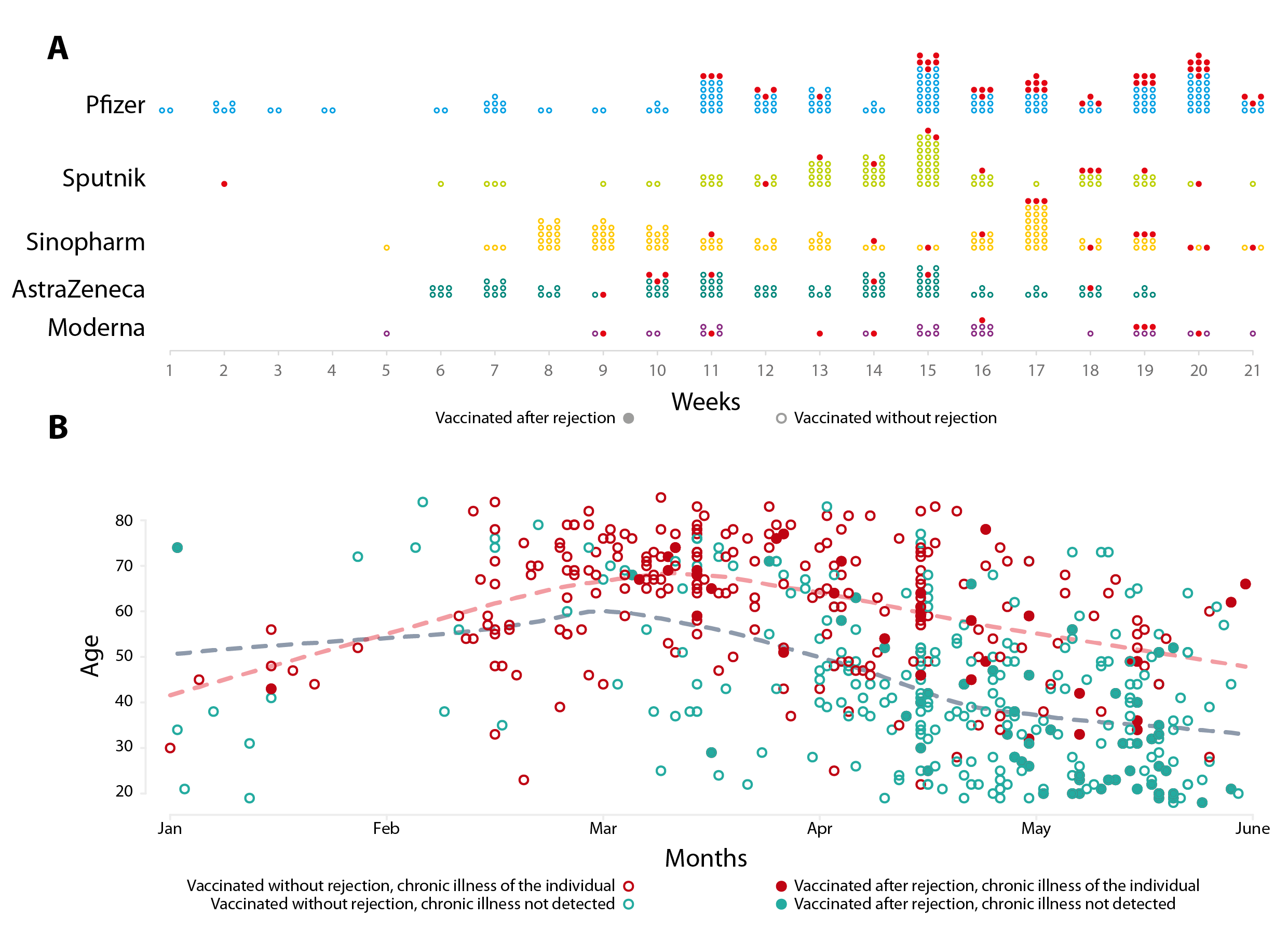}
\caption{\textbf{Dynamics of vaccination by vaccine types and re-selection.} \textbf{A} Dynamics of vaccination by vaccine types. Dots represent number of vaccines taken by weeks. Dark dots represent those vaccines that were taken after rejecting another one. \textbf{B} The dynamics of vaccination by age and chronic illness. The days are identified by first vaccine taken. Those who took a vaccine after rejecting another are depicted with filled markers and those who took the first assigned vaccine are depicted with hollow marker. Dashed lines represent the local averages of vaccinated patients' age (dark is used for patients with chronic illness and bright is used for healthy).}
\label{fig:Fig2}
\end{figure}

Figure \ref{fig:Fig2}A demonstrates the dynamics of revealed preferences by vaccines types. Although Pfizer was available early on, individual selection was only possible later in spring when the prioritized groups has all been already vaccinated. Sinopharm could be selected the same time as Pfizer and it was specifically allocated to the older population. AstraZeneca, Moderna and Sputnik could be chosen relatively early compared to the other vaccines. Figure \ref{fig:Fig2}B shows that rejection and acceptance happened in two cohorts. Elder patients with chronic illness rejected vaccines early so that they could get vaccinated with a more preferred vaccine type in March. The younger individuals received their first vaccine later and could also select a preferred vaccine. Supporting Information 1 contains further description of the dynamics of vaccination and rejection by age, chronic illness and vaccine types.

\subsection{Trusted information sources characterize groups of vaccine preferences}

Let us turn now to the determinants of vaccine hesitancy. Individuals can be influenced in a very complex manner by the public discourse on vaccines. The competing messages of actors can be amplified by the communication channel used (eg. social media) and the mechanisms are difficult to disentangle. To better understand how external influence relates to hesitancy, the respondents were asked to identify their trusted information sources on 
COVID-19 vaccination, as well as their socio-demographic, health-related information, detailed history of vaccination and any infections. The following sources of trusted information were listed: doctors, scientists, anti-vaccine propagators, politicians, family, friends, journalists, celebrities. The respondents also had to name their most frequently used information channels on pandemic measures such as web news,	social media, 	press,	radio,	TV,	family,	friends. Major actors communicated differently about the vaccines. Doctors and scientists argued for vaccination\endnote{Recommendation for vaccination strategy of the Hungarian Academy of Sciences:  https://mta.hu/english/recommendation-of-the-hungarian-academy-of-science-on-the-management-of-the-covid-19-epidemic-in-the-short-and-long-term-110630}, whereas anti-vaccine propagators emphasized the non-effectiveness and the side effects of the COVID-19 vaccines\endnote{A summary on anti-vax movement in Hungary: https://www.iribeaconproject.org/our-work-analysis-and-insights/2021-09-16/hungarys-anti-vax-movement-alive-and-kicking}. Some medical doctors warned against vaccines that are not approved by the EU
\endnote{Scientists, doctors slam Hungarian government criticism of western vaccines:  https://www.reuters.com/article/us-health-coronavirus-hungary-vaccines-idUSKBN2CE2C3}. The government communication campaigned for the non-EU-approved vaccines to improve their rate of acceptance by the population \endnote{The Hungarian prime minister is vaccinated with Sinopharm: https://www.euronews.com/2021/02/28/hungary-s-pm-viktor-orban-vaccinated-against-covid-with-chinese-sinopharm-vaccine} but a heated debate around vaccine types has fragmented the society \endnote{Surveys https://hungarytoday.hu/hungarian-vaccination-politics-vaccine-preference-coronavirus-survey-willingness/}. 

We apply a variable selection method to decrease the complexity of the problem that both the source of information and the channel of communication (eg. social media) can influence hesitancy simultaneously. The LASSO model (short for least absolute shrinkage and selection operator) enables us to select the most important variables that have high-enough predictive power (the process is described in Methods). Estimation results are reported in Supporting Information \(2\), Table \ref{tab:lasso}.  We find that communication channels are not important determinants of hesitancy, because all but one (web news) are dropped from the model. We also see that both the demographic characteristics and the sources of trusted information on COVID-19 vaccination are important predictors of the outcome variables. We rely on this result when we select the explanatory variables to the linear probability models.

\begin{figure}[ht]
\centering
\includegraphics[width=\linewidth]{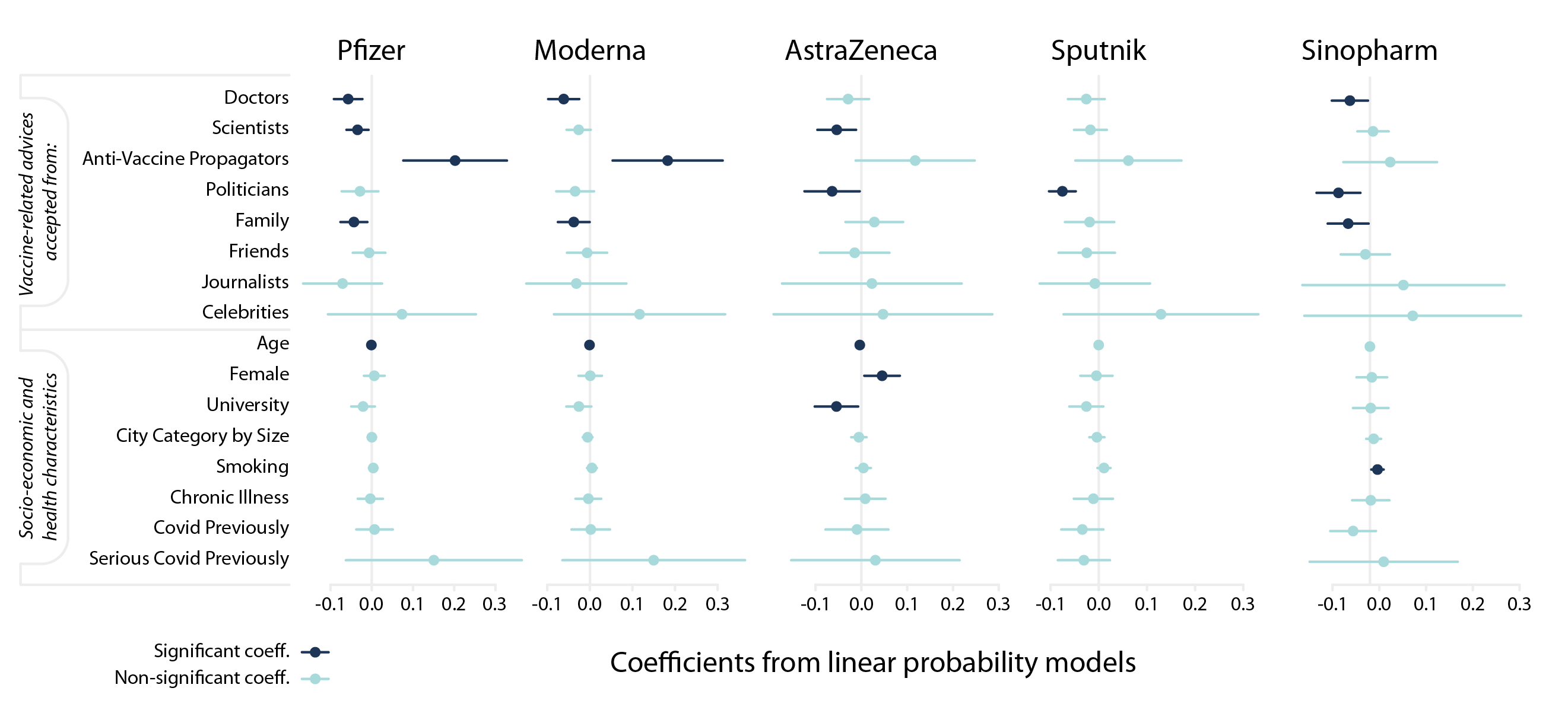}
\caption{\textbf{Source of advice taken characterize hesitancy against vaccines differently.} Estimation results from linear probability models with robust standard errors. The dependent variable equals 1 if the respondent evaluates a vaccine type unacceptable and 0 otherwise. Markers denote point estimates and 95\% confidence intervals are denoted by the bars. Non-significant coefficients are in light blue ($p\geq0.05$); significant coefficients are in dark-blue ($p<0.05$). }
\label{fig:Fig3}
\end{figure}


Next, we estimate linear probability models (LPM) to analyze the factors that shape the evaluation of each vaccine individually. The dependent variables take value of \(1\) if the vaccine is evaluated "unacceptable" and value of \(0\) if it is evaluated either as "acceptable", "fair", "good" or "best". Independent variables include "Advice" indicators that take value 1 if the individual follows the advice of the denoted source. "Age" is a numeric variable and is measured in years, "Female" takes value 1 if the respondent is woman and zero otherwise, "University" equals 1 if the respondent has a university degree or equivalent, "Smoking" and "Chronic Illness" are binary variables as well as "COVID Previously" and "Serious COVID previously". We present the summary statistics in Supporting Information 3, which also contains a correlation matrix and VIF statistics. The latter tests indicate no issues of multicollinearity. The estimation specification can be found in Methods.

Figure \ref{fig:Fig3} illustrates point estimates and $95\%$ confidence intervals calculated from heteroscedasticity-robust standard errors. (For detailed results see Table S3 in Supplementary Information 4.) These results suggest that the factors that affect the rejection decision differ among vaccine types. The most remarkable differences are in the effect of the trusted information sources. Those who trust the advice of anti-vaccine propagators are 18.2 and 20.2 percent more likely to evaluate vaccines made with the newest mRNA technologies (Pfizer and Moderna) unacceptable. On the contrary, trusting doctors' advice decreases the hesitancy probability by 5.7, 6.1, and 5.4 percent for Pfizer, Moderna, and Sinopharm. Trusting scientists reduce the likelihood of evaluating a vaccine unacceptable by 3.5, 2.6 and 5.5 for Pfizer, Moderna, and AstraZeneca. Those who take advice from politicians are 6.4, 7.6 and 8.4 percent more likely to evaluate AstraZeneca, Sputnik and Sinopharm acceptable. Those who trust their family's advice are more likely to evaluate Pfizer, Moderna and Sinopharm acceptable by 4.3, 3.8 and 5.8 percent. The coefficients for the advice of celebrities are either not or only slightly significant. However, the point estimates are large and positive for all types of vaccines. 

Socio-demographic characteristics of respondents are found to correlate with vaccine hesitancy in a plausible way. Age has a negative correlation implying that vulnerable, older individuals are less likely to evaluate vaccines unacceptable. Although the coefficient is small in Figure \ref{fig:Fig3}, the regression on standardized variables reported in Table S3 (in Supporting Information 4) reveals that age is an important factor of evaluating Pfizer, Moderna and AstraZeneca acceptable. Females are more likely to evaluate AstraZeneca unacceptable, probably due to reported cases of females' blood clot\endnote{https://www.reuters.com/business/healthcare-pharmaceuticals/uk-regulator-says-some-evidence-astrazeneca-clots-occur-more-women-than-men-2021-05-06/}. University degree, however, decreases hesitancy against AstraZeneca.

We have carried out various robustness checks that are reported in Supporting Information 4. First, our results hold if we use a logistic regression specification instead of ordinary least squares. Second, our results remain similar after the inclusion of subjective wealth as a control variable. One limitation of our study follows from the fact that the survey was taken when most registered persons have already been vaccinated. This means that "spreading of the alternatives" \cite{luo2017spreading} may have affected our results. That is, after a difficult decision making, the valuation of the accepted alternative increases, whereas, that of the rejected alternative decreases. Our results do not change if we apply the models on a subsample of those who had stable vaccine assessment over the period.


In sum, these results suggest diverse vaccine hesitancy across vaccine types. Vaccines using the newest mRNA technologies (Pfizer and Moderna) are likely to be evaluated unacceptable by those who follow anti-vaccine propagators. The large coefficients imply the effectiveness of anti-vaccine propagators. 
Vaccines using more established vector-based and whole virus technologies are more likely to be evaluated acceptable by those who follow local politicians. This suggests that public communication can make less popular vaccines acceptable for many. We find that the advice of doctors and scientists can mitigate hesitancy against all types of vaccines. Nevertheless, the coefficients of advice taken from doctors, scientists, and politicians are much smaller, half to one-third of the anti-vaccine voices. This difference could explain the forms of communications: the anti-vaccine propagators usually target fears and emotions, whereas doctors, scientists, and politicians tend to use rational arguments. 

\subsection{Individual selection improves vaccine assessment}

\begin{figure}[!b]
\centering
\includegraphics[width=\linewidth]{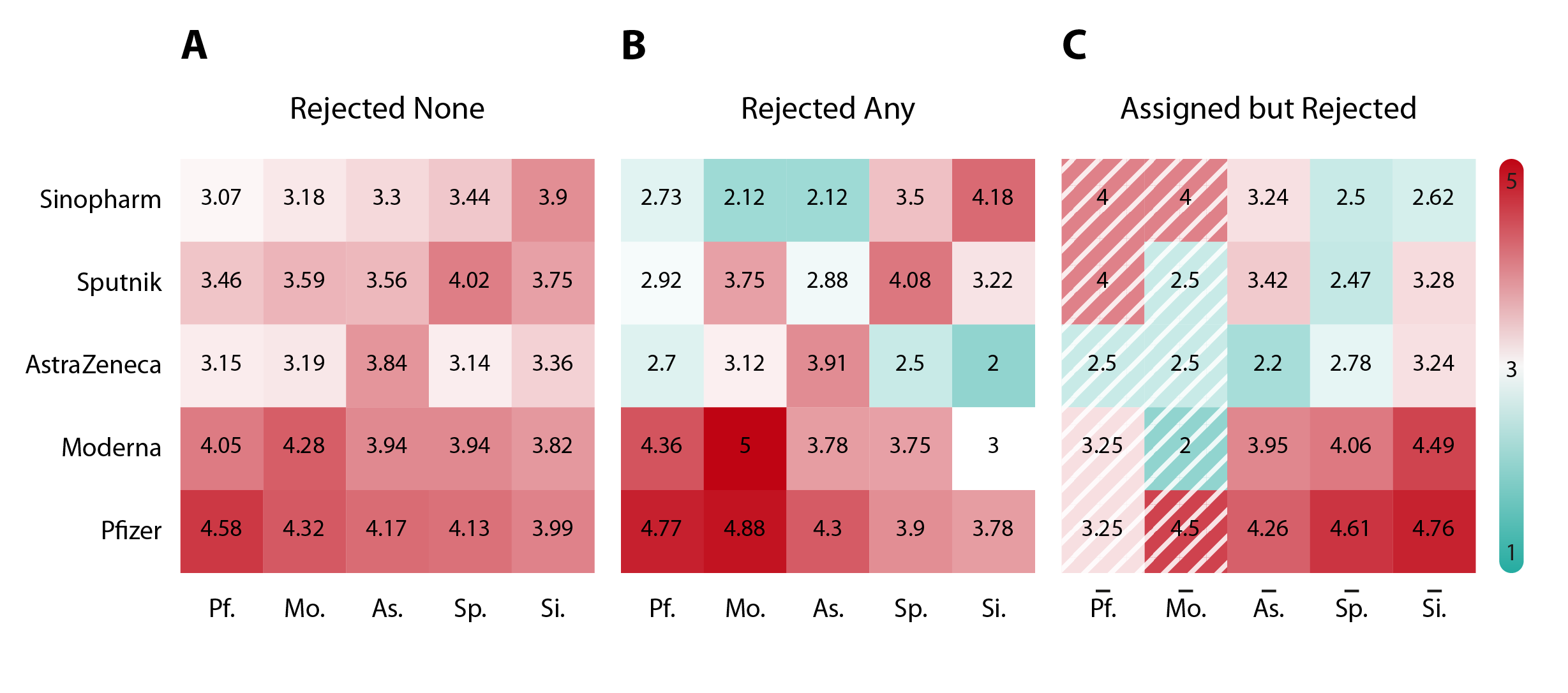}
\caption[respectively.]{\textbf{Average vaccine rating by assigned, re-selected and rejected vaccines.}
\textbf{A} Ratings of those patients who accepted their first vaccine offer. Individuals are grouped by columns of received vaccine. Rows correspond to average rating of a certain vaccine. \textbf{B} Ratings of those patients who rejected at least one assigned vaccine. Individuals are grouped by columns of received vaccine. Rows correspond to average rating of a certain vaccine. \textbf{C} Ratings of those patients who rejected at least one assigned vaccine. Individuals are grouped by columns of rejected vaccine (marked by overline). Rows correspond to average rating of a certain vaccine. The average rating in the first two shaded columns are calculated from few observations only.}
\label{fig:Fig4}
\end{figure}

In this section, we analyze vaccine assessment of patients sorted by received and rejected vaccines. This enables us to inform policy on how free choice of vaccines can impact vaccine acceptance. We can also group vaccines based on the tendency that individuals find both of them acceptable. 

The nationally representative survey asked respondents to rate each of the vaccines based on their personal opinion on a scale from \(1\) to \(5\). 
Figure \ref{fig:Fig4} summarizes the vaccine evaluations in the form of a heat map. A heat map cell corresponds to the average rating of a type of vaccine denoted by the row among a group of patients who got and did not reject any vaccine (Figure \ref{fig:Fig4}A), got and rejected any vaccine (Figure \ref{fig:Fig4}B), or were first offered and rejected (Figure \ref{fig:Fig4}C) the same vaccine denoted by the column. Supporting Information 5 contains summary statistics of vaccine assessments depicted in Figure \ref{fig:Fig4}. 

In Figure \ref{fig:Fig4}A, we consider patients who accepted their first vaccine offer. Notice that Pfizer has the highest average ratings across all columns. This implies that free vaccine choice does not necessarily mean that all patients get the most preferred vaccine. Instead, individuals make an uncertain decision considering acceptable alternatives and 
unknown waiting times between the sequential vaccine offers. In the Supporting Information, Table \ref{tableS12} presents the distribution of the most preferred vaccine when the accepted vaccine is not the most highly-rated vaccine. We infer that patients were willing to accept a vaccine other than the preferred Pfizer or Moderna to avoid the long anticipated waiting times.
In Table \ref{tableS13}, we show that patients had to wait almost \(2.76\) weeks on average to get their most preferred vaccine if they rejected a less preferred vaccine. In Table \ref{tableS14}, we also present evidence that the anticipated waiting times closely follow the actual waiting times on average. That is, if a patient accepts the first vaccine offer then the following conditions must hold; the patient prefers the accepted vaccine to the other vaccines or she has a belief that a more preferred vaccine, which is Pfizer and Moderna with high likelihood, will not be offered anytime soon. We provide an extended analysis in the Supporting Information \(6\).

Figure \ref{fig:Fig4}B illustrates the ratings of those patients who rejected at least one vaccine. We see that the diagonal values in Figure \ref{fig:Fig4}B are slightly higher than the ones in Figure \ref{fig:Fig4}A. This implies that 
individuals who rejected any vaccines before value their received vaccine slightly better than those who accepted their first vaccine. Figure \ref{fig:Fig4}B also shows a high level of dislike towards AstraZeneca, Sinopharm and somewhat Sputnik when it comes to rating vaccines other than the accepted vaccine. 

Comparing the mean values on Figure \ref{fig:Fig4}A and \ref{fig:Fig4}B, we can see fundamental differences between the
rejected none and rejected any groups. Based on the values on the diagonal, those patients who rejected at least one vaccine rated the obtained vaccine higher than those who accepted the first vaccine offer. Looking at the off-diagonal, we see that the rejected any group tends to rate the non-accepted vaccines to be worse. For instance, the mean ratings for AstraZeneca at the off-diagonal are strictly smaller among the patients who rejected a vaccine at least once than among those who accepted their first offer. We also see those patients who received Sinopharm after rejection tend to rate other vaccines worse than those who accepted Sinopharm in their first offer. 

Figure \ref{fig:Fig4}C highlights the benefits of free choice. We group patients based on the first vaccine offer rejection denoted by the columns. We implicitly assume here that the vaccine registration is not affected by the government's decision to allow free choice. Let us point out the number of observations is small for the Pfizer and Moderna groups with only \(6\) and \(2\) observations, respectively. In other words, patients rarely reject Pfizer or Moderna when these vaccines are offered for the first time. Focusing on the AstraZeneca, Sputnik and Sinopharm columns, we see that the values for the assigned vaccines on the diagonal are strictly smaller than any of alternatives. We can also see a strong preference towards Pfizer and Moderna from the AstraZeneca, the Sputnik and the Sinopharm columns. Figure \ref{fig:Fig4}C suggests that free choice significantly increased the efficiency of vaccine allocation because this avoided unnecessary risks of final rejections by enabling re-selecting still acceptable vaccines. 

Assessment of vaccines might have dynamically changed as we learned their effects on health antibody production. We report how assessment of vaccines have changed over the third wave by vaccine types in Supporting Information 6. Most of these stay well below 5 percent of the total sample. Only in case of the AstraZeneca did the number of negative changes almost reach 10 percent of the sample, which is most likely due to the strong negative influence of the news about the side effects. 

Finally, Supporting Information 7 contains a preference analysis of those individuals who were not vaccinated but evaluated minimum 2 but maximum 3 of the investigated vaccines acceptable. This enables us to eliminate those who cannot accept a specific vaccine to look at co-acceptance among vaccines. We find in Figure \ref{fig:SI7} that Pfizer is accepted by many who also accept Moderna (this is the most common co-acceptance link), Sputnik and AstraZeneca, but less so with Sinopharm. Yet, the Sputnik-Sinopharm co-acceptance link is the most frequent one that does not include Pfizer. This suggests that a wide and diverse vaccine portfolio including, both the newest technologies and the older techniques can provide good second-best choices in a society of divergent vaccine preferences.

In sum, we find that free choice across COVID-19 vaccine types allows individuals to get an "acceptable" vaccine and reject the unacceptable alternatives. Although mRNA vaccines are the most preferred ones in our sample, some individuals have specifically chosen vector-based and whole virus vaccines. Our results suggest that the society is fragmented into groups of individuals that either prefers the newest vaccine technologies and refuse the more established vaccine technologies or vice versa. 

\section{Discussion}

The COVID-19 virus created an unprecedented challenge is human history. This is the first global pandemic in the digital age, which concerned all countries around the globe. 
Nations and international alliances, such as the European Union, started financing the development of vaccines in a diverse way, since the fast and reliable supply of effective vaccines seemed the only solution to end the pandemic. A large number of vaccine developments have started simultaneously, led by the producers of the main geopolitical powers (USA, EU, China, Russia and India). So far, several of these vaccines have entered the mass production phase but these use multiple technologies including inactivated whole viruses (eg. Sinopharm), adeno-virus vectors (eg. AstraZeneca, Janssen, Sputnik) and the mRNA technology (eg. Pfizer, Moderna). Most governments decided to use multiple types of vaccines in order to vaccinate the population as soon as possible.

There is a growing consensus that the vaccine diversity, generated by this extraordinary competition between vaccine developers, has several values. First of all, vast capacities have been developed with unprecedented speed when the outcomes - the efficiency of the vaccine - were still very uncertain \cite{andreadakis2020covid, price2020knowledge}. Many have also stressed the need for a production at full and increasing capacities; while others highlight that lower efficacy vaccines are better to use than wait for a more efficient one \cite{castillo2021market}. Furthermore, individual vaccine preferences may differ by technology and country of origin that can be better handled with a diverse vaccine portfolio \cite{schwarzinger2021covid}. This latter aspect raises the question whether governments should
give the freedom of choice to the people over what type of vaccine they wish to get \cite{hughes2021opinion}.

During the third wave of pandemic in Europe in the first half of 2021, which was also the first period of vaccination under vaccine shortage, most European countries gave no choice in between the available vaccines for the registered people when getting their shots. However scientific studies argued the potential benefits of free choice in the UK \cite{dal2021let}. Survey tests in Germany also suggested strong preferences in between the two most different vaccines available there in April, Pfizer and AstraZeneca, providing experimental evidence that free choice can mitigate vaccine hesitancy \cite{sprengholz2021power}. A recent survey conducted in eight European countries shows the differences in vaccine hesitancy and also the divergent preferences of the people over the available vaccines \cite{steinert2021covid}. Our paper adds to this line of research by analyzing a representative survey in Hungary that not just ask the preferences of the patients, but collects also information about their actual decisions over accepting or rejecting offers for specific vaccines, thus providing their revealed preferences over the available vaccines.

The available vaccines differ in many characteristics, such as technology, country of origin, effectiveness, side effects, approval status by national and international authorities, etc. These aspects are very hard to measure, weight and compare. Even if experts are conducting the ranking based on sophisticated techniques
, the result can be different \cite{abdelwahab2021novel} depending on the subjective valuations of the experts. It is then not a surprise that the average patients can have very diverse preferences, that can significantly vary across countries \cite{steinert2021covid}, and can also change over time \cite{schwarzinger2021covid}, as more scientific information becomes available on the effectiveness of the vaccines in the long run and for new mutations, and more precise estimations on the risky side-effects for different groups of populations. Our survey shows surprisingly diverse preferences, where one of the most relevant aspect identified was the technology, namely mRNA versus vector-vaccine and whole-virus. The other major fragmentation in the preferences was in between over the EU-approved Western vaccines (Pfizer, Moderna, AztraZeneca and Janssen) and the Eastern vaccines (Sputnik and Sinopharm) that were additional vaccine types to the common EU purchase.

Besides describing the surveyed preferences and actual choices over the available vaccines, we aimed to understand the reasons behind the preferences. A novel finding of our study, that was not considered in previous international surveys \cite{steinert2021covid}, is the effect of information source that respondents trust. We found that people trusting in politicians were more likely to accept less popular vaccines, in our case vector-based
(AstraZeneca and Sputnik) or whole-virus vaccines (Sinopharm). Those open to conspiracy theories were more likely to reject mRNA vaccines (Pfizer and Moderna). This shows that the free choice policy in vaccine allocation should be combined with the right communication strategy by the government and other decision makers, where different populations should be targeted with different messages when offered with a range of vaccines.

\section{Methods}

\subsection{Data}

The data used in this paper was collected with a CATI telephone survey on a nationally representative sample of the Hungarian adult (18 years or older) population between May 25-31, 2021. The sample size was $1,000$, which is equal to the conventional sample size for nationally representative samples in the country. Data was collected both over mobile and landline phones. This campaign was a part of a larger data collection effort, which started in March 2020 and it is still ongoing. The Hungarian Data Provider Questionnaire \cite{karsai2020hungary,koltai2021monitoring} continuously collects survey data through online questionnaires \cite{MASZKonline} and is monthly asked on nationally representative phone surveys to follow the behavioural changes of people in terms of their contact dynamics, mobility patterns, self-protection habits, and ask people's opinion related to the COVID-19 pandemic. The sample was selected by a multi-step, proportionally stratified, probabilistic sampling procedure. The sample is representative for the Hungarian population aged 18 or older by gender, age, education and domicile. Minor deviations of the data from population ratios were corrected by iterative proportional weighting after the data collection. After data collection, only the anonymised and hashed data was shared with people involved in the project after signing non-disclosure agreements. The data collection and processing have been evaluated and approved by the Research Ethic Committee of the Medical Research Council of Hungary (ETT TUKEB IV/3073-1/2021/EKU).

\subsection{Availability of Data}
For privacy protection we share only the relevant part of the data, which does not allow the reidentification of participants. We exclude sensitive information such as domicile and labels for categorical responses. The data is available at \url{https://doi.org/10.5281/zenodo.5575586}.

\subsection{Methods}
In the LASSO model, the dependent variable takes values between 1 and 5, where 1 is "unacceptable" and 5 is "the best"). In models 1-5, we estimate evaluation values by vaccine types. Then, in models 6 to 9, the ratings' variance, mean, minimum, and maximum serve as dependent variables. In model 10, the dependent variable is binary taking the value of 1 if ever rejected a vaccine offer and 0 otherwise. The dependent variable is 1 in model 11, if any vaccine is unacceptable. The lasso coefficients (\(\widehat{\beta}_{\lambda }^{L}\))  minimize the following expression: \(\sum_{i=1}^{n}(y_{i}-\beta_{0}-\sum_{j=1}^{p}\beta_{j}x_{ij})^{2}+\lambda \sum_{j=1}^{p}\left | \beta_j \right |\) .

Linear probability models are used to predict rejection of vaccines. These regressions are specified by the formula $P(Y=1|X)=\alpha + \beta X + \epsilon$, where $Y$ equals 1 if the vaccine is rated unacceptable, $\alpha$ is the intercept, $\beta$ denotes point estimates, $X$ stand for a vector of independent variables, and $\epsilon$ is heteroscedasticity-robust standard error.

In the heat-maps generation, we used the same vaccine ratings on a scale from \(1\) to \(5\), as we mentioned above. We clustered the survey responses based on the acceptance decision of the first vaccine offer. In case of rejection, we grouped individuals based on the accepted vaccine and the first rejected vaccine. Then, we used standard summary statistics tools to find the number of observation, mean and the standard deviation for each of the groups. We report the mean ratings in Figure \ref{fig:Fig3} and the rest of the summary statistics in Supporting Information \ref{AppFigure3}.

The link weights of the directed network presented in Figure \ref{fig:Fig1} is generated following the aggregation procedure $W_{ab}=\sum_{a}L_{ab}$, where $L_{ab}$ equals 1 if the individual has rejected vaccine $a$ but later received vaccine $b$. The undirected co-occurrence network presented in Figure \ref{fig:SI7} follows the rule of  $W_{mn}=\sum_{m}L_{mn,m \neq n}$ where $L_{mn}$ equals 1 if the respondent accepts both vaccines $m$ and $n$.

\bibliography{vaccine}

\theendnotes

\section*{Acknowledgements}

The authors acknowledge the help of Szabolcs Tóth-Zs in finalizing the figures. Péter Bir\'o gratefully acknowledges financial support from the Hungarian Academy of Sciences, Momentum Grant (No. LP2021-2). Julia Koltai acknowledges funding from the Premium Postdoctoral Grant of the Hungarian Academy of Sciences. The work of Balazs Lengyel was financially supported by Hungarian National Scientific Fund (OTKA K 138970). Agnes Szabo-Morvai was financially supported by the National Research, Development and Innovation Office (OTKA FK141322)

\section*{Author contributions statement}

J.K., A.Sz., G.R., M.K., P.B. and B.L. conceived the survey data collection, K.K., A.Sz., P.B., B.L. designed the investigation, K.K., A.Sz., and B.L. analyzed the data. All authors wrote and reviewed the manuscript. 

\section{Additional information}

\subsection{Accession codes}
All codes to analyze data have been written in R (version 4.0.5) and are available at \url{https://doi.org/10.5281/zenodo.5575586}.

\subsection{Competing interests} 

All authors claim no conflicting interest.

\clearpage

\clearpage

\renewcommand{\thefigure}{S\arabic{figure}}
\renewcommand{\thetable}{S\arabic{table}} 
\renewcommand{\theequation}{S\arabic{equation}} 
\setcounter{figure}{0}
\setcounter{table}{0}
\setcounter{equation}{0}

\section*{Supplementary Information}

\section*{Supporting Information 1: Detailed information on vaccination dynamics}
\begin{figure}[ht]
\centering
\includegraphics[width=0.99\linewidth]{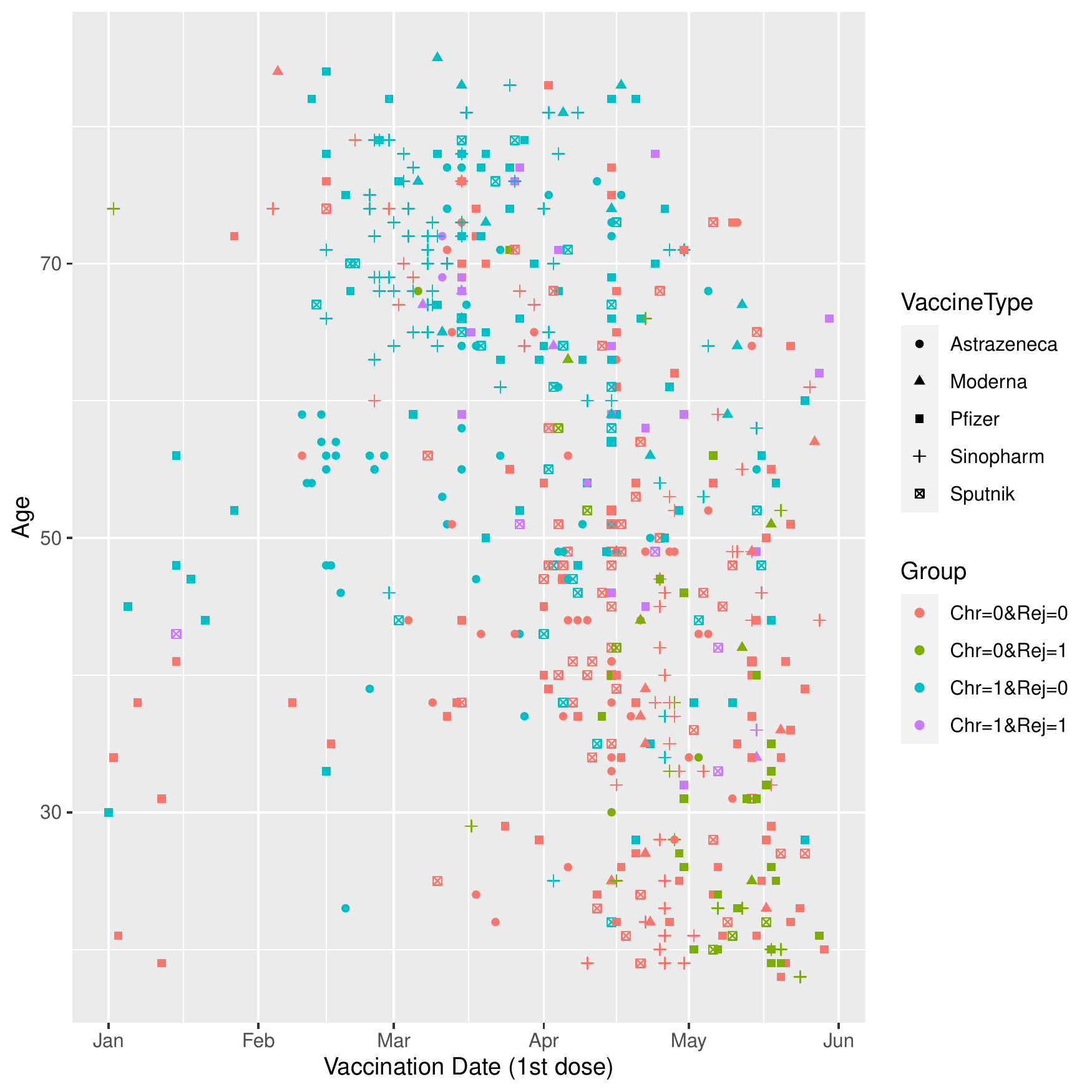}
\caption{Vaccination date, age, vaccination type, chronic illness and rejection status distribution}
\label{fig:Fig5}
\end{figure}

Figure \ref{fig:Fig5} illustrates the time horizon for the vaccination. We can see that Pfizer was the only available vaccine initially, which was most likely offered to medical workers. In the beginning of March, Sinopharm and AstraZeneca became available for mostly chronic ill patients. We see a strict cutoff at the age of \(60\), where Astrazeneca was offered to patients below \(60\) and the Chinese Sinopharm to those who are above \(60.\) Chronic ill patients who rejected a vaccine get vaccinated in the end of March. In April, we see that vaccines become widely available for patients under the age of \(50\). In May, those patients who have rejected a vaccine before and are not chronic ill get finally vaccinated, mainly with Pfizer. Figure \ref{fig:Fig5} suggests that chronic ill and older patients who have not rejected any vaccines in the past tend to get vaccinated sooner.

\clearpage

\begin{figure}[h]
\centering
\includegraphics[width=0.9\linewidth]{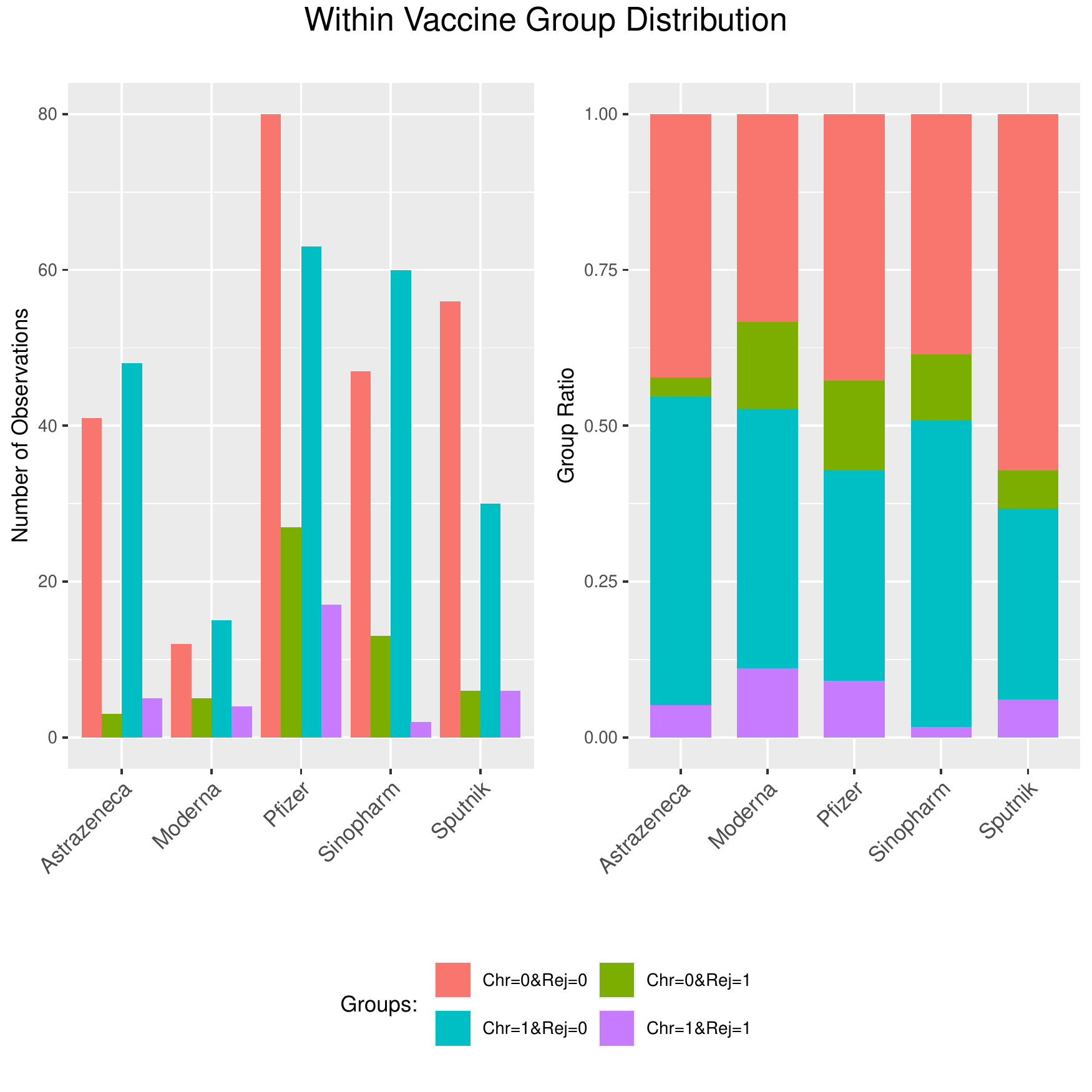}
\caption{Vaccine type distribution conditionally on chronic illness and rejection status.}
\label{fig:FigSI1}
\end{figure}

Figure \ref{fig:FigSI1} shows the distribution of different vaccine types conditionally on chronic illness and rejection status. Out of the \(996\) observations, most patients received Pfizer (\(n=260\)), then Sinopharm (\(n=162\)), Sputnik (\(n=136\)), Astrazeneca (\(n=122\)), Moderna (\(n=45\)) and \(n=261\) people decided not to get vaccinated. We see that only a minority of patients rejected any vaccines. We would like to highlight that Pfizer and Moderna had the highest and AstraZeneca had the lowest ratio of patients who rejected a vaccine before vaccination. We would also like to point out that Sputnik and Pfizer have the two highest not chronic ill to chronic ill ratio of patients. More importantly, Figure \ref{fig:FigSI1} suggests that chronic ill patients are less likely to reject a vaccine than patients without any chronic illness. 

\begin{figure}[h]
\centering
\includegraphics[width=0.9\linewidth]{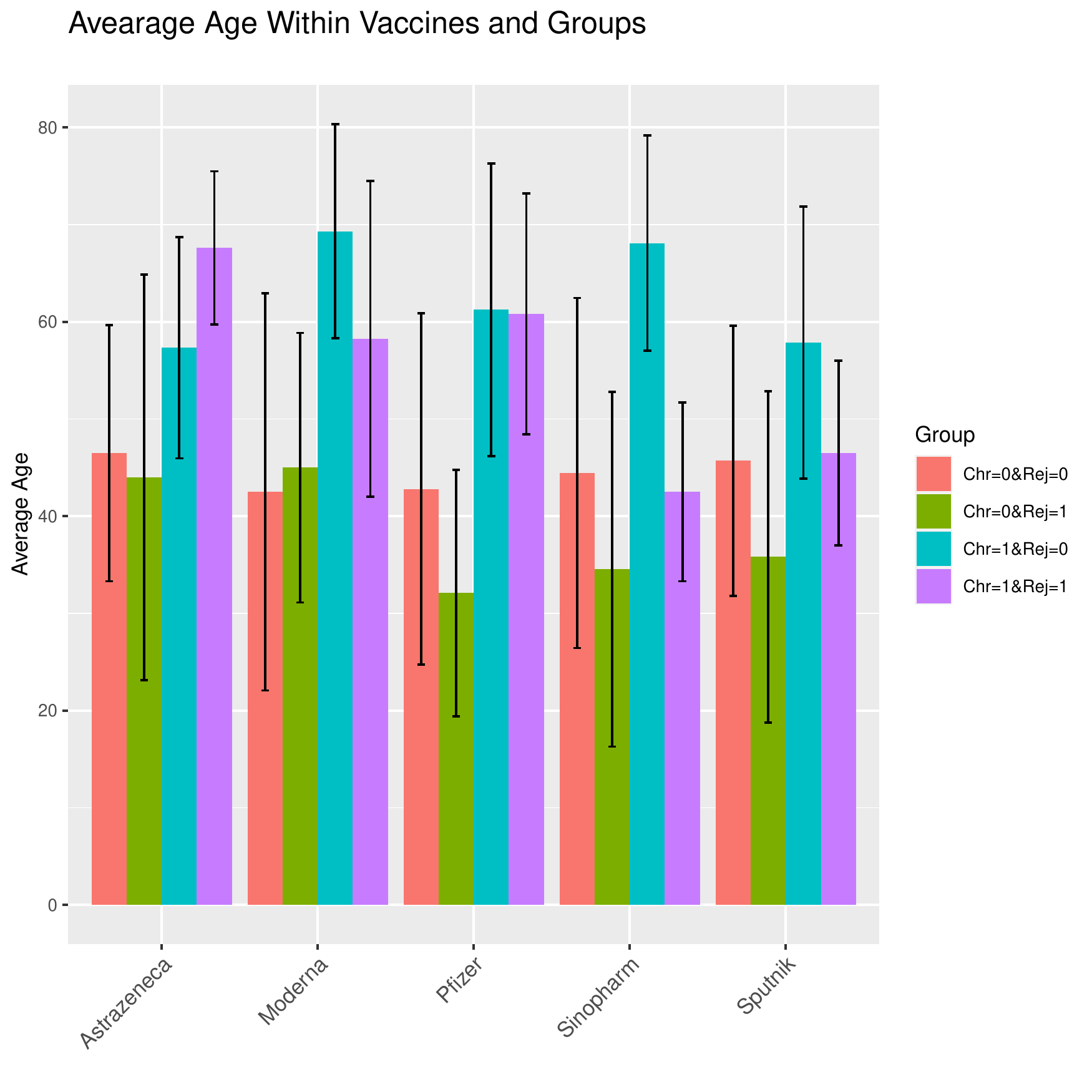}
\caption{Average age distribution conditionally on vaccine type, chronic illness and rejection status. The black line shows \(1\) standard deviation. }
\label{fig:FigSI2}
\end{figure}

\clearpage
Figure \ref{fig:FigSI2} shows the mean and standard deviation of the age distribution conditionally on vaccine type, chronic illness and rejection status. We see that chronic patients tend to be older, as expected. Among patients without chronic illness, Pfizer has the lowest average age, which we believe is a consequence of \(2\) reasons. First, younger medical workers and school teachers could receive Pfizer sooner, as they were prioritized in the vaccine allocation mechanism by the government. Second, younger patients with better physical health felt less threatened by the pandemic and more inclined to wait for their most preferred vaccine, which was Pfizer in most cases. On the other hand, we see that patients without chronic illness, who received AstraZeneca have the highest average age. A potential explanation could be that elderly patient has a stronger time pressure and hence less likely to wait for alternative vaccines. Figure \ref{fig:FigSI2} shows a lot of variation among vaccine types and chronic illness status, which could be explained by prioritization of patients with certain jobs and the amount of time pressure patients experience to wait for an alternative vaccine. 

\begin{figure}[h]
\centering
\includegraphics[width=0.9\linewidth]{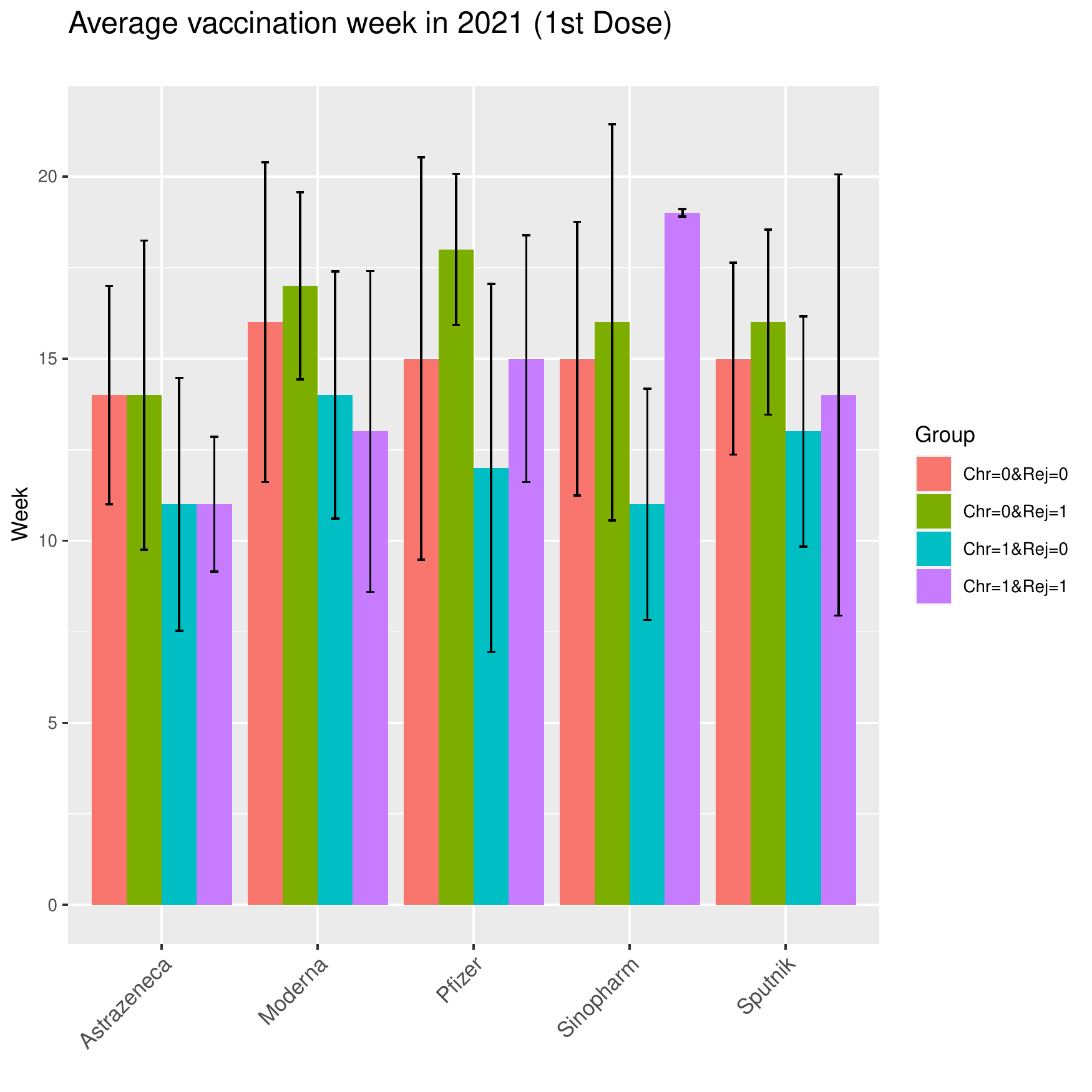}
\caption{Average vaccination week distribution in 2021 conditionally on vaccine type, chronic illness and rejection status. The black line shows \(1\) standard deviation.}
\label{fig:FigSI3}
\end{figure}

\clearpage
Figure \ref{fig:FigSI3} shows the average week of the first vaccination among vaccine types, chronic illness and rejection status. We see that the average vaccination week tends to increase among those who rejected a vaccine regardless of chronic illness status and vaccination type. Let us point out the large variance in the vaccination date among patients who received Pfizer. This indicates that Pfizer was available for the longest time in Hungary. Also, the smaller values for AstraZeneca suggests that patients were more likely to receive AstraZeneca earlier, in February and early March. Figure \ref{fig:FigSI3} is in accordance with Figure \ref{fig:Fig5} as they both show that vaccines became available for the most patient starting from the \(14\)th week and patients with Pfizer have the largest variation in the time of vaccination. 

\clearpage
\section*{Supporting Information 2: Variable selection for the vaccine rejection analysis}

\begin{table}[htbp]
  \centering
  \caption{Lasso models on vaccine rejection}
    \begin{tabular}{lcccccccccc}
          & Pfizer & Moderna & AstraZeneca & Sputnik & Sinopharm & Variance & Mean & Min & Max & Any \\
          & (1)   & (2)   & (3)   & (4)   & (5)   & (6)   & (7)   & (8)   & (9)   & (10) \\
    Age   & x     & x     & x     &       &       &       & x     & x     &       & x \\
    Female &       & x     & x     & x     &       &       &       &       &       &  \\
    City Cat. by Size & x     & x     & x     & x     &       &       &       &       &       &  \\
    University &       & x     &       & x     &       & x     &       & x     &       &  \\
    Smoking &       & x     & x     & x     & x     &       &       &       &       & x \\
    Chronic Illness &       &       & x     & x     &       &       &       & x     &       &  \\
    Acute Illness &       &       &       & x     &       & x     &       &       &       &  \\
    Worked Last Week &       &       &       & x     &       & x     &       & x     &       &  \\
    Wealth Pre COVID-19 &       & x     &       &       &       &       &       &       &       &  \\
    Wealth Now &       & x     & x     & x     & x     & x     & x     & x     &       & x \\
    Adv. from Doct. & x     & x     & x     & x     &       & x     & x     &       & x     & x \\
    Adv. from Pol. &       &       &       & x     &       & x     &       &       &       & x \\
    Adv. from Sci. & x     & x     & x     & x     &       & x     &       & x     & x     & x \\
    Adv. from Anti-vacc. & x     & x     & x     & x     &       & x     &       & x     &       & x \\
    Prop. &      &      &      &      &       &    &       &      &       &  \\
    Adv. from Family &       &       &       &       &       &       &       & x     &       &  \\
    Adv. from Friends &       &       &       & x     &       &       &       &       &       &  \\
    Adv. from Celeb. &       &       &       &       &       & x     &       &       &       &  \\
    Adv. from Journ. &       &       & x     &       &       &       &       & x     &       &  \\
    Online News &       &       & x     & x     & x     & x     &       & x     &       & x \\
    \end{tabular}%
    \begin{tablenotes}
    \item{\textit{Note: Dependent variables: Pfizer and other vaccines: rating (1: "unacceptable"; 5: "best"); Variance/Mean/Min/Max: variance/mean/min/max of the ratings; UnacceptAny=1 if any vaccine rated as "unacceptable". Explanatory variables excluded from each lasso model:	SourceSocialNetwork;	SourcePress;	SourceRadio;	SourceTV;	SourceFamily;	SourceFriends;	SourceOther (VaccineX: Takes advice regarding COVID-19 vaccination from X; SourceX: Source of information used to get to know about pandemic-related government measures)}}
    \end{tablenotes}
  \label{tab:lasso}%
\end{table}%

\clearpage
\section*{Supporting Information 3: Variable description: summary statistics, correlation, VIF}

\begin{table}[ht]
\centering
\caption{Summary statistics and VIF of variables in the linear probability model in Figure \ref{fig:Fig2}}
\begin{tabular}{llllllll}
  \hline
Variable& Min & 1st Qu. & Median & Mean & 3rd Qu. & Max & VIF \\ 
  \hline
Advice from Doctors & 0.000   &  0.000   & 1.000   & 0.729   & 1.000   & 1.000   &  1.121 \\ 
Advice from Scientists & 0.000   & 0.000   & 1.000   & 0.595   & 1.000   & 1.000   & 1.168 \\ 
Advice from Anti-vaccine Propagators & 0.000   & 0.000   & 0.000   & 0.042   & 0.000   & 1.000   &  1.092 \\ 
Advice from Politicians & 0.000   & 0.000   & 0.000   & 0.065   & 0.000   & 1.000   & 1.236 \\ 
Advice from Family & 0.000   & 0.000   & 0.000   & 0.222   & 0.000   & 1.000   &  1.536 \\ 
Advice from Friends & 0.000   & 0.000   & 0.000   & 0.136   & 0.000   & 1.000   &  1.506 \\ 
Advice from Journalists & 0.000   & 0.000   & 0.000   & 0.017   & 0.000   & 1.000   & 1.586 \\ 
Advice from Celebrities & 0.000   & 0.000   & 0.000   & 0.015   & 0.000   & 1.000   &  1.583 \\ 
Age & 18.00   & 34.00   & 48.00   & 48.86   & 64.00   & 85.00   &  1.542 \\ 
Female & 0.00   & 0.00   & 1.00   & 0.53   & 1.00   & 1.00   & 1.075 \\ 
University & 0.000   &.0.000   & 0.000   & 0.217   & 0.000   & 1.000   & 1.181 \\ 
City Cat. by Size & 1.000   & 1.000   & 2.000   & 2.296   & 3.000   & 4.000   & 1.073 \\ 
Wealth Pre COVID-19 & 1.000   & 5.000   & 5.000   & 5.406   & 6.000   & 10.000   & 1.141 \\ 
Smoking & 1.000   & 1.000   & 2.000   & 2.208   & 4.000   & 4.000   & 1.114 \\ 
Chronic Illness & 0.000   & 0.000   & 0.000   & 0.406   & 1.000   & 1.000   &  1.403 \\ 
Covid-19 Previously & 0.000   & 0.000   & 0.000   & 0.095   & 0.000   & 1.000   & 1.157 \\ 
Serious Covid-19 Previously & 0.00   & 0.00   & 0.00   & 0.01   & 0.00   & .00   & 1.123 \\ 
   \hline
\end{tabular}
\end{table} 

\begin{figure}[h]
\centering
\includegraphics[width=0.45\linewidth]{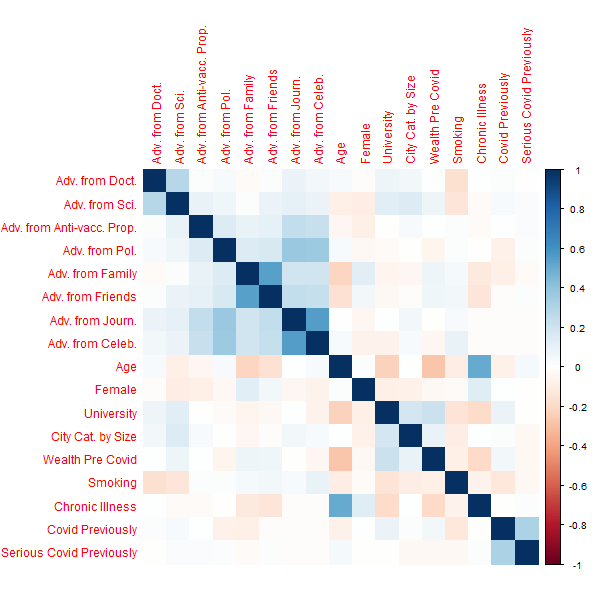}
\caption{Pearson correlation coefficients of variables in the linear probability model in Figure \ref{fig:Fig2}}
\label{fig:correlation}
\end{figure}

\clearpage
\section*{Supporting Information 4: Alternative specifications of the vaccine rejection analysis}

\begin{table}[!htbp] \centering 
  \caption{Linear probability models on evaluating a vaccine unacceptable. Results depicted in Figure \ref{fig:Fig3}} 
  \label{} 
\begin{tabular}{@{\extracolsep{5pt}}lccccc} 
\\[-1.8ex]\hline 
\hline \\[-1.8ex] 
 & \multicolumn{5}{c}{\textit{Dependent variable: Evaluating a vaccine unacceptable}} \\ 
\cline{2-6} 
\\[-1.8ex] & Pfizer & Moderna & Astrazeneca & Sputnik & Sinopharm \\ 
\\[-1.8ex] & (1) & (2) & (3) & (4) & (5)\\ 
\hline \\[-1.8ex] 
 Advice from Doctors & $-$0.057$^{***}$ & $-$0.061$^{***}$ & $-$0.029 & $-$0.026 & $-$0.054$^{**}$ \\ 
  & (0.018) & (0.019) & (0.023) & (0.019) & (0.025) \\ 
 Advice from Scientists & $-$0.035$^{**}$ & $-$0.026$^{*}$ & $-$0.054$^{**}$ & $-$0.017 & 0.008 \\ 
  & (0.014) & (0.014) & (0.022) & (0.017) & (0.021) \\ 
 Advice from Anti-Vaccine Propagators & 0.202$^{***}$ & 0.182$^{***}$ & 0.117$^{*}$ & 0.062 & 0.054 \\ 
  & (0.064) & (0.066) & (0.066) & (0.056) & (0.064) \\ 
 Advice from Politicians & $-$0.029 & $-$0.035 & $-$0.064$^{**}$ & $-$0.076$^{***}$ & $-$0.084$^{***}$ \\ 
  & (0.023) & (0.023) & (0.031) & (0.014) & (0.030) \\ 
 Advice from Family & $-$0.043$^{***}$ & $-$0.038$^{**}$ & 0.028 & $-$0.019 & $-$0.058$^{**}$ \\ 
  & (0.017) & (0.019) & (0.032) & (0.026) & (0.028) \\ 
 Advice from Friends & $-$0.007 & $-$0.007 & $-$0.015 & $-$0.025 & $-$0.012 \\ 
  & (0.020) & (0.024) & (0.039) & (0.030) & (0.034) \\ 
 Advice from Journalists & $-$0.071 & $-$0.032 & 0.023 & $-$0.008 & 0.089 \\ 
  & (0.049) & (0.060) & (0.100) & (0.058) & (0.138) \\ 
 Advice from Celebrities & 0.073 & 0.116 & 0.047 & 0.129 & 0.114 \\ 
  & (0.092) & (0.102) & (0.122) & (0.103) & (0.148) \\ 
 Age & $-$0.001$^{**}$ & $-$0.001$^{**}$ & $-$0.003$^{***}$ & $-$0.0003 & $-$0.0001 \\ 
  & (0.0004) & (0.0005) & (0.001) & (0.001) & (0.001) \\ 
 Female & 0.006 & 0.001 & 0.045$^{**}$ & $-$0.005 & 0.005 \\ 
  & (0.013) & (0.014) & (0.020) & (0.017) & (0.021) \\ 
 University & $-$0.021 & $-$0.026$^{*}$ & $-$0.054$^{**}$ & $-$0.026 & 0.002 \\ 
  & (0.015) & (0.015) & (0.024) & (0.018) & (0.024) \\ 
 City Category by Size & 0.0002 & $-$0.005 & $-$0.005 & $-$0.004 & 0.010 \\ 
  & (0.005) & (0.006) & (0.009) & (0.008) & (0.010) \\ 
 Smoking & 0.004 & 0.005 & 0.004 & 0.011 & 0.020$^{**}$ \\ 
  & (0.005) & (0.006) & (0.008) & (0.007) & (0.008) \\ 
 Chronic Illness & $-$0.004 & $-$0.004 & 0.008 & $-$0.011 & 0.002 \\ 
  & (0.016) & (0.015) & (0.023) & (0.021) & (0.025) \\ 
 COVID-19 Previously & 0.007 & 0.002 & $-$0.010 & $-$0.034 & $-$0.045 \\ 
  & (0.023) & (0.023) & (0.035) & (0.022) & (0.031) \\ 
 Serious COVID-19 Previously & 0.151 & 0.149 & 0.031 & $-$0.031 & 0.037 \\ 
  & (0.109) & (0.109) & (0.094) & (0.028) & (0.101) \\ 
 Constant & 0.146$^{***}$ & 0.166$^{***}$ & 0.314$^{***}$ & 0.120$^{**}$ & 0.095$^{*}$ \\ 
  & (0.038) & (0.041) & (0.054) & (0.047) & (0.055) \\ 
\hline \\[-1.8ex] 
Observations & 999 & 999 & 999 & 999 & 999 \\ 
R$^{2}$ & 0.095 & 0.080 & 0.070 & 0.026 & 0.028 \\ 
Adjusted R$^{2}$ & 0.080 & 0.065 & 0.055 & 0.010 & 0.012 \\ 
\hline 
\hline \\[-1.8ex] 
\textit{Note:}  & \multicolumn{5}{r}{$^{*}$p$<$0.1; $^{**}$p$<$0.05; $^{***}$p$<$0.01} \\ 
\end{tabular} 
\end{table} 

\begin{table}[!htbp] \centering 
  \caption{Linear probability models on evaluating a vaccine unacceptable with standardized variables} 
  \label{} 
\begin{tabular}{@{\extracolsep{5pt}}lccccc} 
\\[-1.8ex]\hline 
\hline \\[-1.8ex] 
 & \multicolumn{5}{c}{\textit{Dependent variable: Evaluating a vaccine unacceptable}} \\ 
\cline{2-6} 
\\[-1.8ex] & Pfizer & Moderna & Astrazeneca & Sputnik & Sinopharm \\ 
\\[-1.8ex] & (1) & (2) & (3) & (4) & (5)\\ 
\hline \\[-1.8ex] 
 Advice from Doctors & $-$0.130$^{***}$ & $-$0.130$^{***}$ & $-$0.041 & $-$0.046 & $-$0.076$^{**}$ \\ 
  & (0.040) & (0.039) & (0.033) & (0.034) & (0.035) \\ 
 Advice from Scientists & $-$0.087$^{**}$ & $-$0.062$^{*}$ & $-$0.085$^{**}$ & $-$0.034 & 0.013 \\ 
  & (0.034) & (0.034) & (0.034) & (0.034) & (0.033) \\ 
 Advice from Anti-Vaccine Propagators & 0.207$^{***}$ & 0.174$^{***}$ & 0.076$^{*}$ & 0.049 & 0.035 \\ 
  & (0.066) & (0.063) & (0.043) & (0.045) & (0.041) \\ 
  Advice from Politicians & $-$0.036 & $-$0.041 & $-$0.051$^{**}$ & $-$0.074$^{***}$ & $-$0.066$^{***}$ \\ 
  & (0.029) & (0.027) & (0.024) & (0.014) & (0.023) \\ 
 Advice from Family & $-$0.092$^{***}$ & $-$0.075$^{**}$ & 0.038 & $-$0.031 & $-$0.077$^{**}$ \\ 
  & (0.035) & (0.038) & (0.043) & (0.043) & (0.037) \\ 
 Advice from Friends & $-$0.011 & $-$0.011 & $-$0.016 & $-$0.034 & $-$0.013 \\ 
  & (0.035) & (0.039) & (0.043) & (0.041) & (0.037) \\ 
 Advice from Journalists & $-$0.047 & $-$0.020 & 0.010 & $-$0.004 & 0.037 \\ 
  & (0.032) & (0.037) & (0.041) & (0.030) & (0.057) \\ 
 Advice from Celebrities & 0.045 & 0.067 & 0.018 & 0.062 & 0.044 \\ 
  & (0.057) & (0.059) & (0.047) & (0.050) & (0.057) \\ 
 Age & $-$0.089$^{**}$ & $-$0.086$^{**}$ & $-$0.200$^{***}$ & $-$0.019 & $-$0.003 \\  & (0.041) & (0.041) & (0.039) & (0.042) & (0.040) \\ 
 Female & 0.016 & 0.002 & 0.072$^{**}$ & $-$0.010 & 0.008 \\ 
  & (0.033) & (0.033) & (0.032) & (0.034) & (0.033) \\ 
 University & $-$0.044 & $-$0.051$^{*}$ & $-$0.072$^{**}$ & $-$0.042 & 0.003 \\ 
  & (0.031) & (0.029) & (0.032) & (0.029) & (0.032) \\ 
 City Category by Size & 0.001 & $-$0.027 & $-$0.018 & $-$0.017 & 0.033 \\ 
  & (0.028) & (0.029) & (0.029) & (0.033) & (0.034) \\ 
 Smoking & 0.023 & 0.028 & 0.018 & 0.054 & 0.081$^{**}$ \\ 
  & (0.035) & (0.035) & (0.034) & (0.035) & (0.034) \\ 
 Chronic Illness & $-$0.009 & $-$0.009 & 0.013 & $-$0.022 & 0.004 \\ 
  & (0.039) & (0.036) & (0.036) & (0.040) & (0.040) \\ 
 COVID-19 Previously & 0.010 & 0.003 & $-$0.009 & $-$0.040 & $-$0.042 \\ 
  & (0.034) & (0.032) & (0.033) & (0.026) & (0.029) \\ 
 Serious COVID-19 Previously & 0.076 & 0.071 & 0.010 & $-$0.012 & 0.012 \\ 
  & (0.055) & (0.052) & (0.030) & (0.011) & (0.032) \\ 
 Constant & 0.0003 & 0.0003 & 0.0001 & 0.0001 & 0.0003 \\ 
  & (0.030) & (0.031) & (0.031) & (0.031) & (0.031) \\ 
 \hline \\[-1.8ex] 
Observations & 999 & 999 & 999 & 999 & 999 \\ 
R$^{2}$ & 0.095 & 0.080 & 0.070 & 0.026 & 0.028 \\ 
Adjusted R$^{2}$ & 0.080 & 0.065 & 0.055 & 0.010 & 0.012 \\ 
\hline 
\hline \\[-1.8ex] 
\textit{Note:}  & \multicolumn{5}{r}{$^{*}$p$<$0.1; $^{**}$p$<$0.05; $^{***}$p$<$0.01} \\ 
\end{tabular} 
\end{table} 

\begin{table}[!htbp] \centering 
  \caption{Logistic regression on evaluating a vaccine unacceptable} 
  \label{} 
\begin{tabular}{@{\extracolsep{5pt}}lccccc} 
\\[-1.8ex]\hline 
\hline \\[-1.8ex] 
 & \multicolumn{5}{c}{\textit{Dependent variable: Evaluating a vaccine unacceptable}} \\ 
\cline{2-6} 
\\[-1.8ex] & Pfizer & Moderna & Astrazeneca & Sputnik & Sinopharm \\ 
\\[-1.8ex] & (1) & (2) & (3) & (4) & (5)\\ 
\hline \\[-1.8ex] 
 Advice from Doctors & $-$1.262$^{***}$ & $-$1.211$^{***}$ & $-$0.298 & $-$0.381 & $-$0.510$^{**}$ \\ 
  & (0.396) & (0.358) & (0.229) & (0.266) & (0.217) \\ 
 Advice from Scientists & $-$1.085$^{**}$ & $-$0.682$^{*}$ & $-$0.573$^{**}$ & $-$0.295 & 0.039 \\ 
  & (0.439) & (0.379) & (0.236) & (0.271) & (0.219) \\ 
 Advice from Anti-Vaccine Propagators & 2.838$^{***}$ & 2.349$^{***}$ & 0.974$^{**}$ & 0.938 & 0.495 \\ 
  & (0.560) & (0.549) & (0.459) & (0.588) & (0.517) \\ 
 Advice from Politicians & $-$1.122 & $-$1.228 & $-$0.939 & $-$3.267$^{***}$ & $-$1.474$^{*}$ \\ 
  & (1.051) & (1.016) & (0.635) & (0.844) & (0.834) \\ 
 Advice from Family & $-$1.441$^{**}$ & $-$1.041$^{*}$ & 0.138 & $-$0.429 & $-$0.722$^{**}$ \\ 
  & (0.629) & (0.554) & (0.278) & (0.441) & (0.363) \\ 
 Advice from Friends & $-$0.479 & $-$0.343 & $-$0.066 & $-$0.461 & $-$0.176 \\ 
  & (0.718) & (0.672) & (0.338) & (0.609) & (0.444) \\ 
 Advice from Journalists & $-$15.071$^{***}$ & $-$0.544 & 0.380 & 0.790 & 1.221 \\ 
  & (1.224) & (1.644) & (1.211) & (0.901) & (1.515) \\ 
 Advice from Celebrities & 3.350$^{***}$ & 2.947$^{***}$ & 0.910 & 2.621$^{**}$ & 1.347 \\ 
  & (0.886) & (0.905) & (1.341) & (1.018) & (1.390) \\ 
 Age & $-$0.029$^{**}$ & $-$0.026$^{**}$ & $-$0.037$^{***}$ & $-$0.004 & $-$0.001 \\ 
  & (0.014) & (0.013) & (0.008) & (0.009) & (0.007) \\ 
 Female & 0.122 & $-$0.006 & 0.467$^{**}$ & $-$0.060 & 0.046 \\ 
  & (0.395) & (0.355) & (0.231) & (0.269) & (0.215) \\ 
 University & $-$0.857 & $-$0.914$^{*}$ & $-$0.525$^{*}$ & $-$0.513 & 0.033 \\ 
  & (0.555) & (0.518) & (0.300) & (0.392) & (0.265) \\ 
 City Category by Size & $-$0.021 & $-$0.179 & $-$0.086 & $-$0.070 & 0.105 \\ 
  & (0.165) & (0.163) & (0.104) & (0.134) & (0.104) \\ 
 Smoking & 0.133 & 0.127 & 0.042 & 0.176$^{*}$ & 0.204$^{**}$ \\ 
  & (0.140) & (0.130) & (0.088) & (0.104) & (0.083) \\ 
 Chronic Illness & $-$0.080 & $-$0.092 & 0.103 & $-$0.203 & 0.017 \\ 
  & (0.481) & (0.409) & (0.276) & (0.338) & (0.262) \\ 
 COVID-19 Previously & 0.097 & 0.005 & $-$0.090 & $-$0.811 & $-$0.558 \\ 
  & (0.610) & (0.592) & (0.394) & (0.667) & (0.459) \\ 
 Serious COVID-19 Previously & 1.941$^{**}$ & 1.727$^{*}$ & 0.321 & $-$13.129$^{***}$ & 0.395 \\ 
  & (0.888) & (0.918) & (1.024) & (0.828) & (1.235) \\ 
 Constant & $-$0.840 & $-$0.557 & $-$0.097 & $-$1.892$^{**}$ & $-$2.295$^{***}$ \\ 
  & (0.933) & (0.850) & (0.550) & (0.741) & (0.582) \\ 
 \hline \\[-1.8ex] 
Observations & 999 & 999 & 999 & 999 & 999 \\ 
Akaike Inf. Crit. & 292.334 & 340.300 & 654.138 & 499.692 & 701.087 \\ 
\hline 
\hline \\[-1.8ex] 
\textit{Note:}  & \multicolumn{5}{r}{$^{*}$p$<$0.1; $^{**}$p$<$0.05; $^{***}$p$<$0.01} \\ 
\end{tabular} 
\end{table} 

\begin{table}[!htbp] \centering 
  \caption{Linear probability models on evaluating a vaccine unacceptable with individual subjective wealth} 
  \label{} 
\begin{tabular}{@{\extracolsep{5pt}}lccccc} 
\\[-1.8ex]\hline 
\hline \\[-1.8ex] 
 & \multicolumn{5}{c}{\textit{Dependent variable: Evaluating a vaccine unacceptable}} \\ 
\cline{2-6} 
\\[-1.8ex] & Pfizer & Moderna & Astrazeneca & Sputnik & Sinopharm \\ 
\\[-1.8ex] & (1) & (2) & (3) & (4) & (5)\\ 
\hline \\[-1.8ex] 
 Advice from Doctors & $-$0.058$^{***}$ & $-$0.062$^{***}$ & $-$0.029 & $-$0.027 & $-$0.054$^{**}$ \\ 
  & (0.018) & (0.019) & (0.023) & (0.019) & (0.025) \\ 
 Advice from Scientists & $-$0.034$^{**}$ & $-$0.026$^{*}$ & $-$0.054$^{**}$ & $-$0.017 & 0.009 \\ 
  & (0.014) & (0.014) & (0.021) & (0.017) & (0.021) \\ 
 Advice from Anti-Vaccine Propagators & 0.202$^{***}$ & 0.182$^{***}$ & 0.117$^{*}$ & 0.061 & 0.054 \\ 
  & (0.064) & (0.066) & (0.066) & (0.056) & (0.064) \\ 
 Advice from Politicians & $-$0.030 & $-$0.037$^{*}$ & $-$0.065$^{**}$ & $-$0.079$^{***}$ & $-$0.088$^{***}$ \\ 
  & (0.022) & (0.022) & (0.030) & (0.015) & (0.030) \\ 
 Advice from Family & $-$0.043$^{***}$ & $-$0.037$^{**}$ & 0.029 & $-$0.018 & $-$0.057$^{**}$ \\ 
  & (0.017) & (0.019) & (0.032) & (0.026) & (0.028) \\ 
 Advice from Friends & $-$0.006 & $-$0.006 & $-$0.014 & $-$0.024 & $-$0.011 \\ 
  & (0.020) & (0.024) & (0.039) & (0.030) & (0.034) \\ 
 Advice from Journalists & $-$0.070 & $-$0.030 & 0.024 & $-$0.006 & 0.092 \\ 
  & (0.049) & (0.060) & (0.100) & (0.058) & (0.138) \\ 
 Advice from Celebrities & 0.071 & 0.114 & 0.046 & 0.124 & 0.109 \\ 
  & (0.091) & (0.102) & (0.122) & (0.100) & (0.145) \\ 
 Age & $-$0.001$^{**}$ & $-$0.001$^{**}$ & $-$0.004$^{***}$ & $-$0.0004 & $-$0.0003 \\ 
  & (0.0005) & (0.0005) & (0.001) & (0.001) & (0.001) \\ 
 Female & 0.006 & 0.0004 & 0.045$^{**}$ & $-$0.005 & 0.004 \\ 
  & (0.013) & (0.014) & (0.020) & (0.017) & (0.021) \\ 
 University & $-$0.019 & $-$0.023 & $-$0.053$^{**}$ & $-$0.021 & 0.008 \\ 
  & (0.014) & (0.015) & (0.024) & (0.018) & (0.025) \\ 
 City Category by Size & 0.001 & $-$0.005 & $-$0.005 & $-$0.003 & 0.011 \\ 
  & (0.005) & (0.006) & (0.009) & (0.008) & (0.010) \\ 
 Wealth pre COVID-19 & $-$0.004 & $-$0.005 & $-$0.003 & $-$0.009$^{*}$ & $-$0.011 \\ 
  & (0.004) & (0.004) & (0.006) & (0.005) & (0.007) \\ 
 Smoking & 0.003 & 0.004 & 0.004 & 0.010 & 0.019$^{**}$ \\ 
  & (0.005) & (0.006) & (0.008) & (0.007) & (0.008) \\ 
 Chronic Illness & $-$0.005 & $-$0.005 & 0.008 & $-$0.013 & $-$0.0002 \\ 
  & (0.016) & (0.015) & (0.023) & (0.021) & (0.025) \\ 
 COVID-19 Previously & 0.008 & 0.003 & $-$0.009 & $-$0.032 & $-$0.043 \\ 
  & (0.023) & (0.023) & (0.035) & (0.022) & (0.031) \\ 
 Serious COVID-19 Previously & 0.148 & 0.146 & 0.029 & $-$0.036 & 0.030 \\ 
  & (0.110) & (0.110) & (0.094) & (0.027) & (0.100) \\ 
 Constant & 0.175$^{***}$ & 0.199$^{***}$ & 0.332$^{***}$ & 0.178$^{***}$ & 0.162$^{**}$ \\ 
  & (0.051) & (0.053) & (0.068) & (0.060) & (0.073) \\ 
\hline \\[-1.8ex] 
Observations & 999 & 999 & 999 & 999 & 999 \\ 
R$^{2}$ & 0.096 & 0.082 & 0.070 & 0.029 & 0.031 \\ 
Adjusted R$^{2}$ & 0.081 & 0.066 & 0.054 & 0.013 & 0.014 \\ 
\hline 
\hline \\[-1.8ex] 
\textit{Note:}  & \multicolumn{5}{r}{$^{*}$p$<$0.1; $^{**}$p$<$0.05; $^{***}$p$<$0.01} \\ 
\end{tabular} 
\end{table} 

\begin{table}[!htbp] \centering 
  \caption{Linear probility models on evaluating a vaccine unacceptable among those individuals who had stable vaccine assessment over the period} 
  \label{} 
\begin{tabular}{@{\extracolsep{5pt}}lccccc} 
\\[-1.8ex]\hline 
\hline \\[-1.8ex] 
 & \multicolumn{5}{c}{\textit{Dependent variable: Evaluating a vaccine unacceptable}} \\ 
\cline{2-6} 
\\[-1.8ex] & Pfizer & Moderna & Astrazeneca & Sputnik & Sinopharm \\ 
\\[-1.8ex] & (1) & (2) & (3) & (4) & (5)\\ 
\hline \\[-1.8ex] 
 Advice from Doctors & $-$0.052$^{***}$ & $-$0.052$^{***}$ & $-$0.018 & $-$0.022 & $-$0.038 \\ 
  & (0.018) & (0.018) & (0.023) & (0.020) & (0.024) \\ 
 Advice from Scientists & $-$0.034$^{**}$ & $-$0.025$^{*}$ & $-$0.042$^{*}$ & $-$0.011 & 0.006 \\ 
  & (0.014) & (0.014) & (0.022) & (0.018) & (0.022) \\ 
 Advice from Anti-Vaccine Propagators & 0.175$^{***}$ & 0.143$^{**}$ & 0.121$^{*}$ & 0.040 & 0.012 \\ 
  & (0.065) & (0.064) & (0.073) & (0.055) & (0.061) \\ 
 Advice from Politicians & $-$0.026 & $-$0.031 & $-$0.060$^{**}$ & $-$0.076$^{***}$ & $-$0.081$^{**}$ \\ 
  & (0.024) & (0.022) & (0.027) & (0.014) & (0.031) \\ 
 Advice from Family & $-$0.041$^{**}$ & $-$0.033$^{*}$ & 0.029 & $-$0.014 & $-$0.055$^{**}$ \\ 
  & (0.017) & (0.019) & (0.031) & (0.027) & (0.028) \\ 
 Advice from Friends & $-$0.006 & $-$0.005 & $-$0.051 & $-$0.023 & $-$0.027 \\ 
  & (0.021) & (0.024) & (0.035) & (0.032) & (0.032) \\ 
 Advice from Journalists & $-$0.068 & $-$0.027 & 0.090 & $-$0.008 & 0.119 \\ 
  & (0.050) & (0.059) & (0.089) & (0.058) & (0.162) \\ 
 Advice from Celebrities & 0.082 & 0.120 & $-$0.102 & 0.131 & 0.122 \\ 
  & (0.104) & (0.103) & (0.064) & (0.102) & (0.158) \\ 
 Age & $-$0.001$^{**}$ & $-$0.001$^{*}$ & $-$0.003$^{***}$ & $-$0.0002 & $-$0.0001 \\ 
  & (0.0005) & (0.0005) & (0.001) & (0.001) & (0.001) \\ 
 Female & 0.009 & 0.002 & 0.048$^{**}$ & $-$0.006 & 0.013 \\ 
  & (0.013) & (0.014) & (0.020) & (0.017) & (0.021) \\ 
 University & $-$0.019 & $-$0.022 & $-$0.045$^{*}$ & $-$0.017 & 0.004 \\ 
  & (0.015) & (0.015) & (0.025) & (0.019) & (0.025) \\ 
 City Category by Size & 0.001 & $-$0.005 & $-$0.001 & $-$0.005 & 0.008 \\ 
  & (0.005) & (0.006) & (0.009) & (0.008) & (0.010) \\ 
 Smoking & 0.004 & 0.007 & 0.011 & 0.012$^{*}$ & 0.019$^{**}$ \\ 
  & (0.005) & (0.006) & (0.008) & (0.007) & (0.008) \\ 
 Chronic Illness & $-$0.004 & $-$0.004 & $-$0.0003 & $-$0.005 & $-$0.003 \\ 
  & (0.016) & (0.015) & (0.023) & (0.021) & (0.026) \\ 
 COVID-19 Previously & 0.012 & 0.006 & $-$0.029 & $-$0.032 & $-$0.040 \\ 
  & (0.024) & (0.023) & (0.033) & (0.024) & (0.033) \\ 
 Serious COVID-19 Previously & 0.175 & 0.151 & 0.079 & $-$0.031 & 0.039 \\ 
  & (0.120) & (0.111) & (0.101) & (0.028) & (0.101) \\ 
 Constant & 0.136$^{***}$ & 0.143$^{***}$ & 0.263$^{***}$ & 0.103$^{**}$ & 0.093$^{*}$ \\ 
  & (0.037) & (0.039) & (0.052) & (0.046) & (0.055) \\ 
 \hline \\[-1.8ex] 
Observations & 962 & 983 & 898 & 951 & 946 \\ 
R$^{2}$ & 0.082 & 0.064 & 0.073 & 0.022 & 0.023 \\ 
Adjusted R$^{2}$ & 0.067 & 0.048 & 0.056 & 0.005 & 0.007 \\ 
\hline 
\hline \\[-1.8ex] 
\textit{Note:}  & \multicolumn{5}{r}{$^{*}$p$<$0.1; $^{**}$p$<$0.05; $^{***}$p$<$0.01} \\ 
\end{tabular} 
\end{table} 

\clearpage
\section*{Supporting Information 5: Summary statistics for vaccine assessment matrices}
\label{AppFigure3}

\begin{table}[ht]
\centering
\caption{Summary statistics for Figure \ref{fig:Fig4}\textbf{A}: Accepted first vaccine} 
\begin{tabular}{llrrr}
  \hline
 Accepted Vaccine & Rated Vaccine & Observations & Mean & Standard Deviation \\ 
  \hline
Astrazeneca & Pfizer & 121 & 4.17 & 1.03 \\ 
  AstraZeneca & Moderna & 121 & 3.94 & 1.00 \\ 
  AstraZeneca & AstraZeneca & 121 & 3.84 & 1.00 \\ 
  AstraZeneca & Sputnik & 121 & 3.56 & 1.12 \\ 
  AstraZeneca & Sinopharm & 121 & 3.30 & 1.24 \\ 
  Moderna & Pfizer &  35 & 4.32 & 1.06 \\ 
  Moderna & Moderna &  35 & 4.28 & 0.92 \\ 
  Moderna & AstraZeneca &  35 & 3.19 & 1.04 \\ 
  Moderna & Sputnik &  35 & 3.59 & 1.18 \\ 
  Moderna & Sinopharm &  35 & 3.18 & 1.36 \\ 
  Pfizer & Pfizer & 202 & 4.58 & 0.70 \\ 
  Pfizer & Moderna & 202 & 4.05 & 1.00 \\ 
  Pfizer & AstraZeneca & 202 & 3.15 & 1.17 \\ 
  Pfizer & Sputnik & 202 & 3.46 & 1.19 \\ 
  Pfizer & Sinopharm & 202 & 3.07 & 1.27 \\ 
  Sinopharm & Pfizer & 146 & 3.99 & 1.17 \\ 
  Sinopharm & Moderna & 146 & 3.82 & 1.13 \\ 
  Sinopharm & AstraZeneca & 146 & 3.36 & 1.23 \\ 
  Sinopharm & Sputnik & 146 & 3.75 & 1.10 \\ 
  Sinopharm & Sinopharm & 146 & 3.90 & 1.15 \\ 
  Sputnik & Pfizer & 123 & 4.13 & 1.08 \\ 
  Sputnik & Moderna & 123 & 3.94 & 1.07 \\ 
  Sputnik & AstraZeneca & 123 & 3.14 & 1.09 \\ 
  Sputnik & Sputnik & 123 & 4.02 & 1.05 \\ 
  Sputnik & Sinopharm & 123 & 3.44 & 1.17 \\ 
   \hline
\end{tabular}
\end{table}
\clearpage

\begin{table}[ht]
\centering
  \caption{Summary statistics for Figure \ref{fig:Fig4}\textbf{B}: Rejected at least \(1\) vaccine; Grouped by accepted vaccine} 
\begin{tabular}{llrrr}
  \hline
 Accepted Vaccine & Rated Vaccine & Observations & Mean & Standard Deviation \\ 
  \hline
AstraZeneca & Pfizer &  11 & 4.30 & 1.25 \\ 
  AstraZeneca & Moderna &  11 & 3.78 & 1.20 \\ 
  AstraZeneca & AstraZeneca &  11 & 3.91 & 0.83 \\ 
  AstraZeneca & Sputnik &  11 & 2.88 & 0.83 \\ 
  AstraZeneca & Sinopharm &  11 & 2.12 & 0.83 \\ 
  Moderna & Pfizer &  10 & 4.88 & 0.35 \\ 
  Moderna & Moderna &  10 & 5.00 & 0.00 \\ 
  Moderna & AstraZeneca &  10 & 3.12 & 0.99 \\ 
  Moderna & Sputnik &  10 & 3.75 & 1.04 \\ 
  Moderna & Sinopharm &  10 & 2.12 & 1.36 \\ 
  Pfizer & Pfizer &  58 & 4.77 & 0.54 \\ 
  Pfizer & Moderna &  58 & 4.36 & 0.86 \\ 
  Pfizer & AstraZeneca &  58 & 2.70 & 1.35 \\ 
  Pfizer & Sputnik &  58 & 2.92 & 1.25 \\ 
  Pfizer & Sinopharm &  58 & 2.73 & 1.39 \\ 
  Sinopharm & Pfizer &  16 & 3.78 & 1.20 \\ 
  Sinopharm & Moderna &  16 & 3.00 & 1.00 \\ 
  Sinopharm & AstraZeneca &  16 & 2.00 & 0.87 \\ 
  Sinopharm & Sputnik &  16 & 3.22 & 0.83 \\ 
  Sinopharm & Sinopharm &  16 & 4.18 & 0.87 \\ 
  Sputnik & Pfizer &  13 & 3.90 & 1.10 \\ 
  Sputnik & Moderna &  13 & 3.75 & 1.16 \\ 
  Sputnik & AstraZeneca &  13 & 2.50 & 1.27 \\ 
  Sputnik & Sputnik &  13 & 4.08 & 0.79 \\ 
  Sputnik & Sinopharm &  13 & 3.50 & 1.18 \\ 
   \hline
\end{tabular}
\end{table}

\clearpage

\begin{table}[ht]
\centering
\caption{Summary statistics for Figure \ref{fig:Fig4}\textbf{C}: Rejected at least \(1\) vaccine; Grouped by assigned vaccine} 
\begin{tabular}{llrrr}
  \hline
 Assigned Vaccine & Rated Vaccine & Observations & Mean & Standard Deviation \\ 
  \hline
AstraZeneca & Pfizer &  28 & 4.26 & 1.14 \\ 
 AstraZeneca & Moderna &  28 & 3.95 & 1.12 \\ 
  AstraZeneca & AstraZeneca &  28 & 2.20 & 1.08 \\ 
  AstraZeneca & Sputnik &  28 & 3.42 & 1.10 \\ 
  AstraZeneca & Sinopharm &  28 & 3.24 & 1.51 \\ 
  Moderna & Pfizer &   2 & 4.50 & 0.71 \\ 
  Moderna & Moderna &   2 & 2.00 & 0.00 \\ 
  Moderna & AstraZeneca &   2 & 2.50 & 0.71 \\ 
  Moderna & Sputnik &   2 & 2.50 & 0.71 \\ 
  Moderna & Sinopharm &   2 & 4.00 & 0.00 \\ 
  Pfizer & Pfizer &   6 & 3.25 & 1.71 \\ 
  Pfizer & Moderna &   6 & 3.25 & 1.71 \\ 
  Pfizer & AstraZeneca &   6 & 2.50 & 1.00 \\ 
  Pfizer & Sputnik &   6 & 4.00 & 0.82 \\ 
  Pfizer & Sinopharm &   6 & 4.00 & 1.22 \\ 
  Sinopharm & Pfizer &  44 & 4.76 & 0.43 \\ 
  Sinopharm & Moderna &  44 & 4.49 & 0.70 \\ 
  Sinopharm & AstraZeneca &  44 & 3.24 & 1.35 \\ 
  Sinopharm & Sputnik &  44 & 3.28 & 1.14 \\ 
  Sinopharm & Sinopharm &  44 & 2.62 & 1.10 \\ 
  Sputnik & Pfizer &  26 & 4.61 & 0.72 \\ 
  Sputnik & Moderna &  26 & 4.06 & 1.00 \\ 
  Sputnik & AstraZeneca &  26 & 2.78 & 1.31 \\ 
  Sputnik & Sputnik &  26 & 2.47 & 1.12 \\ 
  Sputnik & Sinopharm &  26 & 2.50 & 1.51 \\ 
   \hline
\end{tabular}
\end{table}

\clearpage
\section*{Supporting Information 6: Dynamics of vaccine assessment}
In the nationally representative survey, individuals are asked whether they have changed their opinion about each of the available vaccine types over the past few months. In Table \ref{tableS11}, we show the number of observations who declared a negative and positive change, respectively. 

\begin{table}[ht]
\centering 
  \caption{Number of respondents who changed their attitudes towards vaccines} 
  \label{tableS11}
\begin{tabular}{rrr}
  \hline
 & Negative change & Positive change \\ 
  \hline
Pfizer &  19 &  18 \\ 
  Moderna &   5 &  11 \\ 
  AstraZeneca &  90 &  10 \\ 
  Sputnik &  17 &  31 \\ 
  Sinopharm &  34 &  19 \\ 
   \hline
\end{tabular}
\end{table}

In Table \ref{tableS12}, we consider those individuals who accepted a vaccine that they did not rate the highest among the available vaccines. Based on the accepted vaccine type, we calculated the ratio of individuals who rate an other vaccine the highest. For instance, among the \(9\) observations who accepted Pfizer and did not rate Pfizer as their most preferred vaccine there were \(2\) individuals, who rate Moderna as the most preferred vaccine, and hence we obtained \(0.22.\) As individuals were allowed to give the highest rating to multiple vaccines, we see that the sum of the "Best" ratings must be weakly greater than \(1\). Let us point out that Astrazeneca group has the most observations, who rate another vaccine the highest, and then Sputnik and Sinopharm follows. We can also see a strong preference toward Pfizer among those who did not accept their most preferred vaccine, which suggests the limited supply of Pfizer in the observed period in Hungary. 

\begin{table}[ht]
  \caption{Distribution of the most preferred vaccine types when the accepted vaccine is not the most preferred.} 
\centering
  \label{tableS12}
\begin{tabular}{lrrrrrr}
  \hline
 Accepted vaccine & Observations & Pfizer & Moderna & Astrazeneca & Sputnik & Sinopharm \\ 
  \hline
Pfizer &   9 & - & 0.22 & 0.11 & 0.78 & 0.44 \\ 
  Moderna &   3 & 1.00 & - & 0.00 & 0.33 & 0.00 \\ 
  Astrazeneca &  40 & 0.88 & 0.25 & - & 0.10 & 0.10 \\ 
  Sputnik &  29 & 0.93 & 0.48 & 0.07 & - & 0.10 \\ 
  Sinopharm &  29 & 0.79 & 0.41 & 0.17 & 0.17 & - \\ 
   \hline
\end{tabular}
\end{table}

The representative survey asks individuals to estimate the number of weeks that they would be willing to wait to get their most preferred vaccine. In Table \ref{tableS13}, we grouped individuals based on whether they rejected any vaccines and they rate the accepted vaccine the highest. We see that those who accepted a vaccine that they do not rate the highest estimated the expected number of weeks until the best vaccine to be \(2.34\) and \(2.57\) weeks on average if not rejected any and rejected at least \(1\) vaccine, respectively. We see that these estimations are relatively close to \(2.76,\) which is the mean of the actual number weeks that the those individuals waited, who rejected a vaccine and ended up accepting a vaccine that they rate the highest.

\begin{table}[ht]
  \caption{Expected and actual waiting time distribution based on rejecting any vaccines and rating the accepted vaccine the highest.} 
  \label{tableS13}
\centering
\begin{tabular}{rrrrrrr}
  \hline
 Rejected any & Highest rating & Observations & Mean expected & Mean actual & \(90\%\) CI expected & \(90\%\) CI actual \\ 
  & & & (weeks) & (weeks) & (weeks) & (weeks) \\ 
  \hline
0 & 0 & 101 & 2.34 & - & (1.90,2.78) & (-,-)  \\ 
0 & 1 & 526 & 0.29 & - & (0.21,0.36) & (-,-) \\ 
1 & 0 & 9 & 2.57 & 0.83 & (0.58,4.56) & (0.11,1.56) \\ 
1 & 1 & 99 & 0.25 & 2.76 & (0.11,0.39) & (2.20,3.32) \\ 
   \hline
\end{tabular}
\end{table}

In Table \ref{tableS14}, we grouped individuals, who accepted a vaccine that they do not rate the highest and individuals who were first offered and rejected that vaccine, and ended up getting a vaccine that they rate the highest. First, notice that the waiting times are not homogeneous among the vaccine types. For instance, the mean expected and the mean actual waiting times for Sinopharm are \(3.46\) and \(3.78\) weeks, respectively, while for Pfizer we have only \(1.20\) and \(1.00\). Let us also highlight the trend that AstraZeneca has a higher acceptance to rejection (\(\frac{40}{25}\)) ratio than the one that Sputnik or Sinopharm has (\(\frac{29}{23}\) and \(\frac{29}{41}\), respectively). This could be explained by that the AstraZeneca group estimated the expected waiting time to be longer than it actually was, while we see the opposite for Sputnik and Sinopharm.

\begin{table}[ht]
  \caption{Expected and actual waiting times for accepting and rejecting a vaccine that is not the most preferred.}
  \label{tableS14}
\centering
\begin{tabular}{lrrrrrr}
  \hline
 Vaccine type & Observations & Observations & Mean expected & Mean actual & \(90\%\) CI expected & \(90\%\) CI actual \\ 
  & Accept & Reject & Accept & Reject & Accept & Reject \\
  &  &  & (week) & (week) & (week) & (week) \\
  \hline
AstraZeneca & 40 & 25 & 2.61 & 1.94 & (1.87,3.36) & (1.47,2.42) \\ 
Moderna & 3 & 1 & 0.00 & 2.00 & (-,-) & (-,-)  \\ 
Pfizer & 9 & 5 & 1.20 & 1.00 & (0.09,2.31) & (-0.35,2.35) \\ 
Sinopharm & 29 & 41 & 3.46 & 3.78 & (2.37,4.55) & (2.60,4.96) \\ 
Sputnik & 29 & 23 & 1.61 & 2.58 & (1.04,2.19) & (1.37,3.80) \\ 
   \hline
\end{tabular}
\end{table}

\clearpage
\section*{Supporting Information 7: Co-acceptance of vaccines among non-vaccinated}


\begin{figure}[h]
\centering
\includegraphics[width=0.6\linewidth]{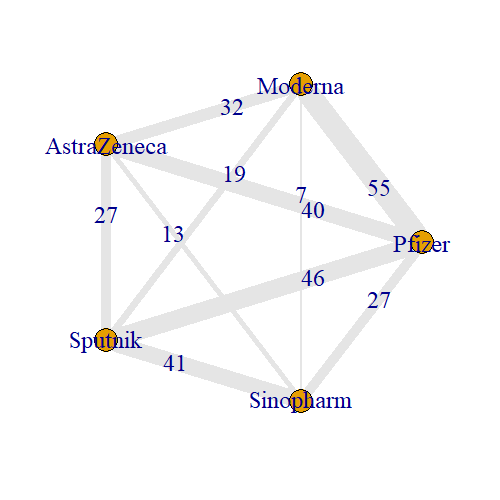}
\caption{Co-acceptance of vaccines. Edge labels reflect to the number of individuals who accept both vaccines included. We only include those individuals to create this network who accept 2 or 3 vaccines.}
\label{fig:SI7}
\end{figure}

\end{document}